\documentclass{IEEEtran}
\usepackage[mathscr]{eucal}
\usepackage[cmex10]{amsmath}
\usepackage{epsfig,epsf,psfrag}
\usepackage{amssymb,amsmath,amsthm,amsfonts,latexsym}
\usepackage{amsmath,graphicx,bm,xcolor,url,overpic}
\usepackage{fixltx2e}
\usepackage{array}
\usepackage{verbatim}
\usepackage{bm}
\usepackage{algorithmic}
\usepackage{algorithm}
\usepackage{verbatim}
\usepackage{textcomp}
\usepackage{mathrsfs}
\usepackage{epstopdf}

\newcommand{\openone}{\leavevmode\hbox{\small1\normalsize\kern-.33em1}}

\catcode`~=11 \def\UrlSpecials{\do\~{\kern -.15em\lower .7ex\hbox{~}\kern .04em}} \catcode`~=13 

\allowdisplaybreaks[4]

\newcommand{\nn}{\nonumber}

\newcommand{\calA}{\mathcal{A}}
\newcommand{\calB}{\mathcal{B}}
\newcommand{\calC}{\mathcal{C}}

\newcommand{\calJ}{\mathcal{J}}

\newcommand{\calP}{\mathcal{P}}
\newcommand{\calQ}{\mathcal{Q}}
\newcommand{\calR}{\mathcal{R}}

\newcommand{\calU}{\mathcal{U}}
\newcommand{\calV}{\mathcal{V}}
\newcommand{\calW}{\mathcal{W}}
\newcommand{\calX}{\mathcal{X}}
\newcommand{\calY}{\mathcal{Y}}
\newcommand{\calZ}{\mathcal{Z}}

\newcommand{\bJ}{\mathbf{J}}

\newcommand{\rmc}{\mathrm{c}}

\newcommand{\rmH}{\mathrm{H}}

\newcommand{\rmQ}{\mathrm{Q}}

\newcommand{\rmT}{\mathrm{T}}

\newcommand{\rmV}{\mathrm{V}}

\newcommand{\bbN}{\mathbb{N}}

\newcommand{\bbR}{\mathbb{R}}

\DeclareMathAlphabet{\mathbsf}{OT1}{cmss}{bx}{n}
\DeclareMathAlphabet{\mathssf}{OT1}{cmss}{m}{sl}

\DeclareSymbolFont{bsfletters}{OT1}{cmss}{bx}{n}  
\DeclareSymbolFont{ssfletters}{OT1}{cmss}{m}{n}
\DeclareMathSymbol{\bsfGamma}{0}{bsfletters}{'000}
\DeclareMathSymbol{\ssfGamma}{0}{ssfletters}{'000}
\DeclareMathSymbol{\bsfDelta}{0}{bsfletters}{'001}
\DeclareMathSymbol{\ssfDelta}{0}{ssfletters}{'001}
\DeclareMathSymbol{\bsfTheta}{0}{bsfletters}{'002}
\DeclareMathSymbol{\ssfTheta}{0}{ssfletters}{'002}
\DeclareMathSymbol{\bsfLambda}{0}{bsfletters}{'003}
\DeclareMathSymbol{\ssfLambda}{0}{ssfletters}{'003}
\DeclareMathSymbol{\bsfXi}{0}{bsfletters}{'004}
\DeclareMathSymbol{\ssfXi}{0}{ssfletters}{'004}
\DeclareMathSymbol{\bsfPi}{0}{bsfletters}{'005}
\DeclareMathSymbol{\ssfPi}{0}{ssfletters}{'005}
\DeclareMathSymbol{\bsfSigma}{0}{bsfletters}{'006}
\DeclareMathSymbol{\ssfSigma}{0}{ssfletters}{'006}
\DeclareMathSymbol{\bsfUpsilon}{0}{bsfletters}{'007}
\DeclareMathSymbol{\ssfUpsilon}{0}{ssfletters}{'007}
\DeclareMathSymbol{\bsfPhi}{0}{bsfletters}{'010}
\DeclareMathSymbol{\ssfPhi}{0}{ssfletters}{'010}
\DeclareMathSymbol{\bsfPsi}{0}{bsfletters}{'011}
\DeclareMathSymbol{\ssfPsi}{0}{ssfletters}{'011}
\DeclareMathSymbol{\bsfOmega}{0}{bsfletters}{'012}
\DeclareMathSymbol{\ssfOmega}{0}{ssfletters}{'012}

\newcommand{\hatE}{\hat{E}}

\newcommand{\tilU}{\tilde{U}}

\newcommand{\tilV}{\tilde{V}}

\newcommand{\tilW}{\tilde{W}}

\newcommand{\tilX}{\tilde{X}}

\newcommand{\tilY}{\tilde{Y}}

\newcommand{\tilZ}{\tilde{Z}}

\newcommand{\barE}{\bar{E}}

\newcommand{\barL}{\bar{L}}

\newcommand{\barQ}{\bar{Q}}
\newcommand{\barR}{\bar{R}}

\newtheorem{theorem}{Theorem} 
\newtheorem{lemma}[theorem]{Lemma}

\newtheorem{definition}{Definition}

\usepackage{tikz,pgfplots,picture,graphicx}
\usepackage{amsmath}
\usepackage{amssymb}
\usepackage{bbm,cite}
\usepackage{color}

\usetikzlibrary{intersections}
\usepgfplotslibrary{fillbetween}
\usepackage[ colorlinks = true,
            linkcolor = blue,
            urlcolor  = blue,
            citecolor = red,
            anchorcolor = green,]{hyperref}
\usepackage{enumitem}
\begin{document}

\title{Privacy-Utility Tradeoff for Hypothesis Testing Over A Noisy Channel}
\author{
Lin Zhou and Daming Cao

\thanks{Lin Zhou is with the School of Cyber Science and Technology, Beihang University, Beijing 100191, China, and also with the Beijing Laboratory for General Aviation Technology, Beihang University, Beijing 100191, China (Email: lzhou@buaa.edu.cn).}
\thanks{Daming Cao is with the School of Computing at the National University of Singapore (dcscaod@nus.edu.sg).}
}

\maketitle

\begin{abstract}
We study a hypothesis testing problem with a privacy constraint over a noisy channel and derive the performance of optimal tests under the Neyman-Pearson criterion. The fundamental limit of interest is the privacy-utility tradeoff (PUT) between the exponent of the type-II error probability and the leakage of the information source subject to a constant constraint on the type-I error probability. We provide an exact characterization of the asymptotic PUT for any non-vanishing type-I error probability. Our result implies that tolerating a larger type-I error probability cannot improve the PUT. Such a result is known as a strong converse or strong impossibility theorem. To prove the strong converse theorem, we apply the recently proposed technique in (Tyagi and Watanabe, 2020) and further demonstrate its generality. The strong converse theorems for several problems, such as hypothesis testing against independence over a noisy channel (Sreekumar and G\"und\"uz, 2020) and hypothesis testing with communication and privacy constraints (Gilani \emph{et al.}, 2020), are established or recovered as special cases of our result.
\end{abstract}

\begin{IEEEkeywords}
Strong converse, information leakage, noisy channel, non-asymptotic converse, Euclidean information theory
\end{IEEEkeywords}

\section{Introduction}

In the binary hypothesis testing problem, given a test sequence $X^n$ and two distributions $P$ and $Q$, one is asked to determine whether the test sequence $X^n$ is generated i.i.d. from $P$ or $Q$. The performance of any test is evaluated by the tradeoff between the type-I and type-II error probabilities. Under the Neyman Pearson setting where the type-I error probability is upper bounded by a constant, the likelihood ratio test~\cite{poor2013introduction} is proved optimal. Chernoff-Stein lemma~\cite{chernoff1952measure} states that the type-II error probability decays exponentially fast with exponent $D(Q\|P)$ when the type-I error probability is upper bounded by one half and the length of the test sequence tends to infinity. This result was later refined by by Strassen~\cite{strassen1962asymptotische} who provided exact second-order asymptotic characterization of the type-II error exponent for any non-vanishing type-I error probability. Strassen's result implies the asymptotic type-II error exponent remains $D(Q\|P)$ regardless of the non-vanishing type-I error probability. Such a result is known as a strong converse theorem, which implies that tolerating a larger type-I error probability cannot increase the asymptotic decay rate of the type-II error probability of an optimal test.

The simple binary hypothesis testing problem was later generalized to various scenarios. Motivated by the application where the source sequence might be only available to a decision maker via rate-limited communication, Ahlswede and Csisz\'ar~\cite{ahlswede1986hypothesis} initiated the study of the hypothesis testing problem with communication constraints. The authors of \cite{ahlswede1986hypothesis} gave exact asymptotic characterization of the rate-exponent tradeoff subject to a vanishing type-I error probability and proved a strong converse result for the special case of testing against independence. Recently, motivated by the fact the source sequence is transmitted over a noisy channel in certain applications, e.g., in a sensor network~\cite{Sohraby07}, Sreekumar and G\"und\"uz~\cite{sreekumar2019distributed} further generalized the setting of \cite{ahlswede1986hypothesis} by adding a noisy channel between the transmitter and the decision maker. However, the authors of \cite{sreekumar2019distributed} derived only a weak converse result which holds for vanishing type-I error probability. For the case of testing against independence, a strong converse result\footnote{The authors of \cite{sreekumar2020strong} claimed that the strong converse result could be extended to other values of $\tau$ in a footnote.} was proved in \cite{sreekumar2020strong} when the bandwidth expansion ratio $\tau$ (defined as the ratio between the number of channel uses $n$ and the length of the source sequence $k$) is $1$ by combining the blowing up lemma~\cite{csiszar2011information} and the strong converse technique recently proposed in~\cite{tyagi2020strong}.

Another generalization of the binary hypothesis testing framework takes \emph{privacy} into consideration. Privacy gains increasing attention from all parties. Releasing collected raw data for statistical inference can potentially leak critical information of individuals (cf. \cite[Fig. 1]{sankar2013utility}). Motivated by the privacy concerns in modern data analyses and machine learning, Liao \emph{et al.} \cite{liao2017hypothesis} applied a privacy mechanism to the original sequences to remove private parts and then studied the hypothesis testing problem with a privacy constraint. In particular, the authors of \cite{liao2017hypothesis} derived the privacy-utility tradeoff~\cite{sankar2013utility} between the decay rate of the type-II error probabilitity and the leakage of the information sources measured with mutual information. Subsequently, the setting in \cite{liao2017hypothesis} was generalized to the case under the maximal leakage privacy constraint in \cite{liao2017hypothesis2} and to the case with communication constraints by Gilani \emph{et al.}~\cite{gilani2019distributed}.

\begin{figure*}[tb]
\centering
\setlength{\unitlength}{0.5cm}
\scalebox{0.7}{
\begin{picture}(35,5)
\linethickness{1pt}
\put(0.5,0.8){\makebox{$U^k$}}
\put(2,1){\vector(1,0){2}}
\put(4,0){\framebox(6,2){Privacy Mechanism}}
\put(10,1){\vector(1,0){3}}
\put(11.5,1.5){\makebox(0,0){$Z^k$}}
\put(13,0){\framebox(4,2){Transmitter}}
\put(17,1){\vector(1,0){3}}
\put(18.5,1.5){\makebox(0,0){$X^n$}}
\put(20,0){\framebox(5,2){Noisy Channel}}
\put(25,1){\vector(1,0){3}}
\put(26.5,1.5){\makebox(0,0){$Y^n$}}
\put(28,0){\framebox(4,2){Detector}}
\put(30,4){\vector(0,-1){2}}
\put(30.2,4.5){\makebox(0,0){$V^k$}}
\put(32,1){\vector(1,0){3}}
\put(33.5,1.5){\makebox(0,0){$\rmH_1$}}
\put(33.5,0.5){\makebox(0,0){$\rmH_2$}}
\end{picture}}
\caption{Hypothesis testing over a noisy channel with a privacy constraint. The transmitter observes source information $U^k$ and applies a privacy mechanism to obtain non-private information $Z^k$. Subsequently, the transmitter encodes $Z^k$ into a codeword $X^n$, which is passed through a noisy channel to yield the output $Y^n$. Given $Y^n$ and side information $V^k$, the detector decides between two hypotheses on the generating distribution of $(U^k,V^k)$. The problem of interest is the privacy-utility tradeoff between the transmitter and the detector, which refers to a tradeoff between the privacy of source information $U^k$ and the error probability of the binary hypothesis test at the detector.}
\label{systemmodel}
\end{figure*}
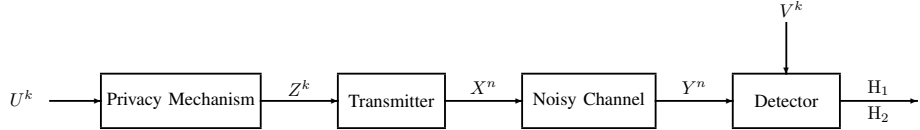

Motivated by i) practical applications where there is a noisy channel between the detector and the transmitter for a hypothesis test and ii) the privacy concerns of statistical data inference problems, we study the privacy-utility tradeoff for a generalized model of \cite{sreekumar2018privacy,gilani2019distributed}. In particular, we consider a hypothesis testing problem over a noisy channel with a privacy constraint as shown in Figure \ref{systemmodel}. We use mutual information as the privacy measure, which is consistent with existing literature in terms of measuring privacy~\cite{liao2017hypothesis2,sankar2013utility,zhou2020multiple,ye2005isit,gilani2019distributed} or security~\cite{bloch2011physical,csiszar2004secrecy,maurer1993}. Such a formulation is intuitive since a small value of the mutual information between two random variables implies a low dependence level. The extreme case of vanishing mutual information privacy constraint, a.k.a. the high privacy limit, ensures almost perfect privacy where virtually no information about the raw data is disclosed. Furthermore, a mutual information constraint can also be motivated by the communication rate constraint as in distributed detection~\cite{viswanathan1997distributed}.

There are many other privacy measures, such as the maximal leakage~\cite{liao2017hypothesis2}, the distortion function~\cite{sreekumar2018privacy}, the differential privacy~\cite{dwork2008differential} and the maximal $\alpha$-leakage~\cite{liao2018tunable}. Among all these privacy measures, the different privacy is probably the most popular one and finds wide applications in various domains~\cite{dwork2011firm}. But a differentially private mechanism might have high leakage under the mutual information privacy measure~\cite{du2012privacy}. Furthermore, the authors of \cite{makhdoumi2014information} show that the expected information leakage under any privacy measure can be upper bounded by a function of the mutual information privacy constraint. Motivated by the results in~\cite{du2012privacy,makhdoumi2014information} and consistent with pioneering works~\cite{sankar2013utility,liao2017hypothesis}, we choose mutual information as the privacy measure in this paper. It is of definite interest to generalize our results to other privacy measures and compare the privacy-utility tradeoffs under different privacy measures as done in \cite{kalantari2018robust}.

Our main contribution is the exact characterization of the privacy-utility tradeoff (PUT) between the decay rate of type-II error probability and the information leakage at the transmitter subject to a constraint on the type-I error probability. It turns out that the exact PUT is a non-convex optimization problem, which can not be solved efficiently. Under the high privacy limit, we derive an easily computable approximation to the PUT using the Euclidean information theory~\cite{borade2008,huang2015euclidean}. Euclidean information theory is based on the local approximation of the KL divergence $D(P\|Q)$ using Taylor expansions when two distributions $P$ and $Q$ are close to each other.

The rest of the paper is organized as follows. In Section \ref{sec:pf}, we formally set up the notation, formulate the hypothesis testing problem with a privacy constraint over a noisy channel and define the privacy-utility tradeoff. Subsequently, we present our characterization of the PUT in Section \ref{sec:main}. The proofs of our results are given in Section \ref{sec:proof}. Finally, in Section \ref{sec:conc}, we conclude the paper and discuss future research directions. The proofs of all supporting lemmas are deferred to appendices.

\section{Problem Formulation}
\label{sec:pf}
\subsection*{Notation}
Random variables and their realizations are in upper (e.g.,\ $X$) and lower case (e.g.,\ $x$) respectively. All sets are denoted in calligraphic font (e.g.,\ $\calX$). We use $\calX^{\rmc}$ to denote the complement of $\calX$. Let $X^n:=(X_1,\ldots,X_n)$ be a random vector of length $n$ and $x^n$ its realization. All logarithms are base $e$. We use $\bbR_+$ and $\bbN$ to denote the set of non-negative real numbers and  natural numbers,  respectively. Given any positive integer $a\in\bbN$, we use $[a]$ to denote $\{1,\cdots, a\}$. The set of all probability distributions on a finite set $\calX$ is denoted as $\calP(\calX)$. For quantities such as entropy and mutual information, we follow the notation in \cite{csiszar2011information}. In particular, when the joint distribution of $(X,Y)$ is $P_{XY}\in\calP(\calX\times\calY)$, we use $I(X;Y)$ and $I(P_X,P_{Y|X})$ interchangeably.

\subsection{Problem Setting}
Let $\calU,\calV,\calZ$ be three finite alphabets and let $P_{UV}$ and $Q_{UV}$ be two probability mass functions defined on the alphabet $\calU\times\calV$. Consider a discrete memoryless channel $P_{Y|X}$ where the input alphabet $\calX$ and the output alphabet $\calY$ are both finite. As we shall see, $\calU$ and $\calZ$ are the input and output alphabets of a privacy mechanism, respectively. We do not impose constraints on the relationship between $|\calU|$ and $|\calZ|$ because our results hold for any finite sets $\calU,\calZ$ and in some cases, the output alphabet $\calZ$ of a privacy mechanism could be larger than the input alphabet $\calU$, e.g., due to noise addition.

We consider the hypothesis testing problem with a privacy constraint in Figure \ref{systemmodel}. A source sequence $U^k$ is observed at the transmitter and another sequence $V^k$ is observed at the detector. In order to infer the relationship between two observations, the transmitter sends a message over the memoryless channel $P_{Y|X}$ to the receiver. Given the transmitted messages, the decoder then checks whether $V^k$ is jointly distributed with $U^k$ according to $P_{UV}$ or $Q_{UV}$ via a binary hypothesis test. For the sake of privacy, the transmitter first applies a privacy mechanism $P_{Z^k|U^k}$ to $U^k$ and obtains non-private information $Z^k$. Subsequently, a function of $Z^k$, known as a message, is transmitted to the receiver over the noisy channel $P_{Y|X}$.

We study the case of testing against independence, i.e., $Q_{UV}=P_UP_V$ where $P_U$ and $P_V$ are induced marginal distributions of $P_{UV}$. We are interested in optimal communication protocols and privacy mechanisms to achieve two goals: i) guarantee the privacy constraint for $U^k$ at the transmitter and ii) ensure reliable decision at the detector. These two goals compete with each other and naturally introduce a privacy-utility tradeoff. Our main results provide exact characterization of the PUT in the asymptotic setting.

Formally, a communication protocol is defined as follows.
\begin{definition}
A communication protocol $(f^{n,k},g^{n,k})$ with $n$ channel uses for hypothesis testing against independence over a noisy channel consists of
\begin{enumerate}
\item a potentially stochastic encoder $f^{n,k}:\calZ^k\to\calX^n$
\item a decoder $g^{n,k}:\calY^n\times\calV^k\to\{\rmH_1,\rmH_2\}$
where 
\begin{itemize}
\item $\rmH_1$: the sequences $U^k$ and $V^k$ are correlated, i.e., $(U^k,V^k)\sim P_{UV}^k$
\item $\rmH_2:$ the sequences $U^k$ and $V^k$ are independent, i.e., $(U^k,V^k)\sim P_U^kP_V^k$.
\end{itemize}
\end{enumerate}
\end{definition}
When $f^{n,k}$ is a stochastic encoder, we use $P_{f^{n,k}}(x^n|z^k)$ to denote the probability that the output of the encoder is $x^n$ when the input is $z^k$. In particular, when $f^{n,k}$ is deterministic, $P_{f^{n,k}}(x^n|z^k)$ is simply an indicator function and outputs $1$ if and only if $x^n=f^{n,k}(z^k)$. Given any communication protocol $(f^{n,k},g^{n,k})$ and any privacy mechanism $P_{Z^k|U^k}$, their performance is evaluated by the type-I and type-II error probabilities:
\begin{align}
\beta_1(f^{n,k},g^{n,k})
&:=\Pr\{g^{n,k}(Y^n,V^k)=\rmH_2|\rmH_1\}\label{err:type1},\\*
\beta_2(f^{n,k},g^{n,k})
&:=\Pr\{g^{n,k}(Y^n,V^k)=\rmH_1|\rmH_2\}\label{err:type2},
\end{align}
where $Y^n$ is the output of passing $X^n=f^{n,k}(Z^k)$ over the noisy memoryless channel $P_{Y|X}$. Thus, the probability terms in the right-hand side of \eqref{err:type1} and \eqref{err:type2} depend on the encoding function $f^{n,k}$ and the privacy mechanism $P_{Z^k|U^k}$ implicitly via the noisy output $Y^n$.

\subsection{Definition of the Privacy-Utility Tradeoff}
We restrict ourselves to memoryless privacy mechanisms, i.e., $P_{Z^k|U^k}=P_{Z|U}^k$ for some $P_{Z|U}\in\calP(\calZ|\calU)$. In fact, the adoption of a memoryless privacy mechanism is consistent with a large body of existing literature~\cite{liao2017hypothesis2,liao2017hypothesis,gilani2019distributed,sreekumar2018privacy,lau2020ICCASP}. Furthermore, the memoryless privacy scheme enjoys low complexity and is motivated by the case where each respondent can apply the same randomized privacy mechanism before submitting replies to queries. In contrast, if one adopts a non-memoryless privacy mechanism, then as the length $k$ of the source sequence increases, one needs to design a different privacy mechanism and suffers from higher complexity, especially in the case of large $k$. Finally, adopting a memoryless privacy mechanism does not trivialize the problem. In fact, our proof, especially the converse proof in Section \ref{sec:proof}, requires us to judiciously combine the analyses for the utility and the privacy.

Under the Neyman-Pearson formulation, we are interested in the maximal type-II error exponent subject to a constant constraint on the type-I error probability $\varepsilon\in(0,1)$, a bandwidth expansion ratio $\tau\in\bbR_+$ and a privacy constraint $L\in\bbR_+$ for $n\in\bbN$ channel uses, i.e.,
\begin{align}
\nn&E^*(k,\tau,L,\varepsilon)\\*
\nn&:=\sup\{E\in\bbR_+:~\exists~(f^{n,k},g^{n,k},P_{Z|U})~\mathrm{s.t.}~n\leq k\tau\\*
\nn&\qquad\qquad I(P_U,P_{Z|U})\leq L,~\beta_1(f^{n,k},g^{n,k})\leq \varepsilon\\*
&\qquad\qquad\beta_2(f^{n,k},g^{n,k})\leq \exp(-kE)\}\label{def:put}.
\end{align}
The privacy constraint $I(P_U,P_{Z|U})\leq L$ implies that the leakage of information source $U^k$ from the privatized version $Z^k$ satisfies that $I(U^k;Z^k)\leq kL$. Note that the privacy is measured using mutual information~\cite{cover2012elements}. 
Such a choice of the privacy measure is consistent with most literature studying physical layer security, e.g,~\cite{gilani2019distributed,liao2017hypothesis,zhou2020multiple,maurer1993,ye2005isit}. 

We remark that $E^*(k,\tau,L,\varepsilon)$ represents a tension between the privacy and the utility. Evidently, the looser the privacy constraint $L$, the better the utility $E^*(k,\tau,L,\varepsilon)$. In the extreme case of $L\geq H(U)$, our setting reduces to the case without a privacy constraint as in \cite[Theorem 2]{sreekumar2019distributed} and achieves the best utility. In the other extreme of $L=0$, we achieve the perfect privacy while the utility $E^*(k,\tau,L,\varepsilon)=0$. This is because to ensure perfect privacy, we generate a private sequence $Z^k$, which is independent of the source sequence $U^k$, and therefore, even the full knowledge of $Z^k$ provide no information about the correlation with side information $V^k$, let alone noisy observations of $Z^k$. To better understand the privacy-utility tradeoff for non-extremal values of $L$, we provide exact characterization of $E^*(k,\tau,L,\varepsilon)$ in the limit of large $k$ for any parameters $(\tau,L,\varepsilon)\in\bbR_+^2\times(0,1)$.

\section{Main Results}
\label{sec:main}
In this section, we present our main results, which exactly characterize the privacy-utility tradeoff in the limit of large $k$. 

\subsection{Achievability}
In this subsection, we present our achievability result, which provides a lower bound on $E^*(k,\tau,L,\varepsilon)$. Several definitions are needed. The capacity~\cite{cover2012elements} of a noisy channel with transition matrix $P_{Y|X}$ is
\begin{align}
C(P_{Y|X})=\max_{P_X\in\calP(\calX)}I(P_X,P_{Y|X})\label{def:capacity}.
\end{align}
Furthermore, let $W$ be an auxiliary random variable taking values in the alphabet $\calW$ and let $\calQ$ denote the set of all joint distributions defined on the alphabet $\calU\times\calV\times\calZ\times\calW$. Given any $P_{Z|U}\in\calP(\calZ|\calU)$, define the following set of distributions 
\begin{align}
\nn&\calQ(P_{UV},P_{Z|U})
:=\{Q_{UVZW}\in\calQ:~Q_{UV}=P_{UV}\\*
&\quad Q_{Z|U}=P_{Z|U},~V-U-Z-W,~|\calW|\leq |\calZ|+1\}
\label{def:calQPC}.
\end{align}

Given any $Q_{UVZW}$, let other distributions denoted by $Q$ be induced distributions. For any $(\tau,L)\in\bbR_+^2$, define the following optimization problem
\begin{align}
\nn&f(\tau,L,P_{UV},P_{Z|U},P_{Y|X})\\*
&:=\max_{\substack{Q_{UVZW}\in\calQ(P_{UV},P_{Z|U}):\\ I(Q_Z,Q_{W|Z})\leq \tau C(P_{Y|X}),~I(Q_U,Q_{Z|U})\leq L}} I(Q_V,Q_{W|V})\label{def:putf}.
\end{align}
Since $V-U-Z-W$ forms a Markov chain under any distribution $Q_{UVZW}\in\calQ(P_{UV},P_{Z|U})$, we have $f(\tau,L,P_{UV},P_{Z|U},P_{Y|X})\leq L$.

Our achievability result states as follows.
\begin{theorem}
\label{weakconverse}
For any $(\tau,L)\in\bbR_+^2, \varepsilon \in(0,1]$,
\begin{align}
\lim_{k\to\infty}E^*(k,\tau,L,\varepsilon)\geq \max_{P_{Z|U}}f(\tau,L,P_{UV},P_{Z|U},P_{Y|X}).
\end{align}
\end{theorem}
Theorem \ref{weakconverse} is a straightforward extension of \cite[Theorem 2]{sreekumar2019distributed} and thus its proof is omitted. The proof of Theorem \ref{weakconverse} proceeds in three steps. Firstly, we calculate the optimal memoryless privacy scheme $P_{Z|U}^*$. Secondly, we apply the memoryless privacy mechanism $P_{Z|U}^*$ to privatize the original information source $U^k$ and obtain the non-private information counterpart $Z^k$. Finally, we study a hypothesis testing problem against independence over a noisy channel for the new source sequence $Z^k$ and the side information $V^k$ at the decoder. The final step is exactly the same as \cite[Theorem 2]{sreekumar2019distributed} when specialized to the case of testing against independence.

\subsection{Converse and Discussions}
Our main contribution in this paper is the following theorem, which presents a non-asymptotic upper bound on the optimal type-II exponent $E^*(k,\tau,L,\varepsilon)$. 
\subsubsection{Preliminaries}
To present our result, for any $(\lambda_1,\lambda_2)\in\bbR_+^2$, define two constants
\begin{align}
\nn&c(\lambda_1,\lambda_2,\tau)\\*
&:=\log|\calV|+(\lambda_1+\lambda_2)\log|\calZ|+\lambda_1 \tau\log|\calY|,\label{def:cabt}\\
\nn&\zeta(\lambda_1,\lambda_2,\gamma,\tau)\\*
\nn&:=3\sqrt{\frac{2c(\lambda_1,\lambda_2,\tau)}{\gamma}}\bigg(\log\frac{|\calW||\calV|}{\sqrt{\frac{2c(\lambda_1,\lambda_2,\tau)}{\gamma}}}\\*
\nn&+\lambda_1\log\frac{|\calZ||\calW|}{\sqrt{\frac{2c(\lambda_1,\lambda_2,\tau)}{\gamma}}}+\lambda_2\log\frac{|\calU||\calZ|}{\sqrt{\frac{2c(\lambda_1,\lambda_2,\tau)}{\gamma}}}\bigg)\\
&+3\lambda_1 \tau\sqrt{\frac{2c(\lambda_1,\lambda_2,\tau)}{\tau\gamma}}\log\frac{|\calX||\calY|}{\sqrt{\frac{2c(\lambda_1,\lambda_2,\tau)}{\tau\gamma}}}\label{def:cbgt},
\end{align}
where $|\calW|$ is a finite constant. Given any distributions $(Q_{UVZW},Q_{XY})$, for any $(\lambda_1,\lambda_2)\in\bbR_+^2$, define the following linear combination of mutual information terms
\begin{align}
\nn&R_{\lambda_1,\lambda_2}^{\tau,L}(Q_{UVZW},Q_{XY})\\*
\nn&:=I(Q_V,Q_{W|V})-\lambda_1(I(Q_Z,Q_{W|Z})-\tau I(Q_X,Q_{Y|X}))\\*
&\qquad-\lambda_2(I(Q_U,Q_{Z|U})-L)\label{def:RbQ}.
\end{align}
Furthermore, define the following optimization value
\begin{align}
\nn&g_{\lambda_1,\lambda_2}^{\tau,L}(P_{UV},P_{Z|U},P_{Y|X})\\*
&:=\sup_{\substack{Q_{UVZW}\in\calQ(P_{UV},P_{Z|U})\\Q_{XY}\in\calC:Q_{Y|X}=P_{Y|X}}}R_{\lambda_1,\lambda_2}^{\tau,L}(Q_{UVZW},Q_{XY})\label{def:rb},
\end{align}
where $\calC$ denotes the set of all joint distributions defined on the alphabet $\calX\times\calY$. As we shall show, $g_{\lambda_1,\lambda_2}^{\tau,L}(P_{UV},P_{Z|U},P_{Y|X})$ is closely related with $f(\tau,L,P_{UV},P_{Z|U},P_{Y|X})$.

For subsequent analysis, given any $P_{Z|U}$, define the mutual information density
\begin{align}
\imath(u;z|P_{Z|U}):=\log\frac{P_{Z|U}(z|u)}{P_Z(z)},~\forall(u,z)\in\calU\times\calZ\label{def:mi},
\end{align}
where $P_Z$ is induced by $P_U$ and $P_{Z|U}$. Note that $\mathbb{E}_{P_{UZ}}[\imath_{U;Z}]=I(P_U,P_{Z|U})$. Define the variance and the third absolute moment of the information density as
\begin{align}
\rmV(P_{Z|U})&:=\mathrm{Var}_{P_{UZ}}[\imath(U;Z|P_{Z|U})],\\
\rmT(P_{Z|U})&:=\mathbb{E}_{P_{UZ}}\big[\big|\imath(U;Z|P_{Z|U})-I(P_U,P_{Z|U})\big|^3\big].
\end{align}
Finally, given any constant $\varepsilon\in(0,1)$, define $L(P_{Z|U},\varepsilon)$ as in \eqref{def:L} on the top of the next page.
\begin{figure*}
\begin{align}
L(P_{Z|U},\varepsilon)
=\left\{
\begin{array}{ll}
\sqrt{\rmV(P_{Z|U})}\rmQ^{-1}\left(\varepsilon-\frac{\rmT(P_{Z|U})}{6\sqrt{k\rmV(P_{Z|U})^3}}\right)&\mathrm{if~}\rmV(P_{Z|U})>0\\
0&\mathrm{otherwise}
\end{array}
\right.
\label{def:L}.
\end{align}
\hrulefill
\end{figure*}

\subsubsection{Main Result}

Our converse result states as follows.
\begin{theorem}
\label{mainresult}
Given any $\varepsilon\in(0,1)$, for any $(\lambda_1,\lambda_2,\gamma)\in\bbR_+^3$ and any $P_{Z|U}\in\calP(\calZ|\calU)$,
\begin{align}
E^*(k,\tau,L,\varepsilon)
\nn&\leq g_{\lambda_1,\lambda_2}^{\tau,L}(P_{UV},P_{Z|U},P_{Y|X})+\zeta(\lambda_1,\lambda_2,\gamma,\tau)\\*
\nn&\qquad-\frac{(6\lambda_1+3\lambda_2+2\gamma)\log(1-\varepsilon)}{k}\\*
\nn&\qquad+\frac{(9\lambda_1+3\lambda_2+3\gamma)\log 2}{k}\\*
&\qquad+\frac{\lambda_2 L(P_{Z|U},(1-\varepsilon)/4)}{\sqrt{k}}.
\end{align}
Furthermore, the strong converse theorem follows as a corollary, i.e., for any $\varepsilon\in(0,1)$,
\begin{align}
\lim_{k\to\infty}E^*(k,\tau,L,\varepsilon)\leq \max_{P_{Z|U}}f(\tau,L,P_{UV},P_{Z|U},P_{Y|X})\label{strong:converse1}.
\end{align}
\end{theorem}
The proof of Theorem \ref{mainresult} is given in Section \ref{sec:proof}. Our proof is based on the recently proposed strong converse technique by Tyagi and Watanabe~\cite{tyagi2020strong} that uses the change of measure technique and variational formulas~\cite{oohama2015exponent,oohama2016wynerziv}. In particular, we first derive a multiletter upper bound on the privacy-utility tradeoff using the change of measure technique. Subsequently, we single letterize the bound using standard information theoretical techniques~\cite{cover2012elements}. Finally, using the alternative variational characterization of $f(\tau,L,P_{UV},P_{Z|U},P_{Y|X})$ established via the supporting hyperplane, we managed to obtain the desired result in Theorem \ref{mainresult}. Our proof applies the strong converse technique by Tyagi and Watanabe~\cite{tyagi2020strong} to a hypothesis testing problem over a noisy channel with a privacy constraint and thus demonstrates the generality of the technique.

We make several additional remarks. Combining the strong converse result in \eqref{strong:converse1} and Theorem \ref{weakconverse}, we conclude that given any $(L,\tau)\in\bbR_+^2$, for any $\varepsilon\in(0,1)$,
\begin{align}
\lim_{k\to\infty}E^*(k,\tau,L,\varepsilon)
&=\max_{P_{Z|U}}f(\tau,L,P_{UV},P_{Z|U},P_{Y|X})\label{strong:converse}\\
&=:f(\tau,L,P_{UV},P_{Y|X}).
\end{align}
Thus, we provide a complete characterization of the asymptotic privacy-utility tradeoff for hypothesis testing against independence over a noisy channel. Our result implies that the asymptotically optimal PUT is \emph{independent} of the type-I error probability for any given privacy constraint. Therefore, tolerating a larger type-I error probability cannot increase the privacy-utility tradeoff of optimal privacy and communication protocols when the lengths of sequences tend to infinity. Such a result is known as \emph{strong converse} in information theory (cf.~\cite{zhou2016cilossy,liu2016brascamp,wei2009strong}), which refines the classical weak converse argument valid only for vanishing type-I error probability.

Furthermore, since several problems are special cases of our formulation, the result in \eqref{strong:converse} implies strong converse and provides complete asymptotic characterization of fundamental limits for all these special cases,~e.g.,~\cite{sreekumar2019distributed,gilani2019distributed,ahlswede1986hypothesis}. In particular, by letting $L\geq H(P_U)$, our setting reduces to the hypothesis testing problem against independence~(special case of \cite[Theorem 2]{sreekumar2019distributed}). A strong converse theorem was \emph{not} established for any $\tau\neq 1$ prior to our work. If one considers a memoryless channel and imposes a communication constraint, i.e., $P_{Y|X}$ is the identity matrix and $\calX=\calY=\{1,\ldots,M\}$ for some $M\in\bbN$, our setting then reduces to the setting of hypothesis testing with both communication and privacy constraints considered in \cite{gilani2019distributed}. The authors of \cite{gilani2019distributed} proved a strong converse result for their setting using the complicated blowing up lemma idea~\cite{csiszar2011information}. Our result here provides an alternative yet simpler proof for their setting.

Finally, we compare our converse result with existing works on hypothesis testing over a noisy channel or with a privacy constraint, especially \cite{gilani2019distributed} and \cite{sreekumar2020strong}. The former one corresponds to the special case where the channel is noiseless. By considering a noisy channel in this paper, our analysis is more \emph{complicated} since we need to account for additional errors due to the noisy nature of the channel. Our results imply the strong converse result in \cite[Theorem 2]{gilani2019distributed} but not vice versa. In \cite{sreekumar2020strong}, \emph{without} a privacy constraint by letting $L\geq H(U)$, the authors proved a strong converse result by combining the techniques in \cite{tyagi2020strong} and the blowing up lemma~\cite{csiszar2011information}. In contrast, our proof is  more transparent and much simpler by getting rid of the blowing up lemma.

\subsection{Illustration of the PUT via a Numerical Example}
\label{sec:nume}
\begin{figure}[tb]
\centering
\includegraphics[width=.9\columnwidth]{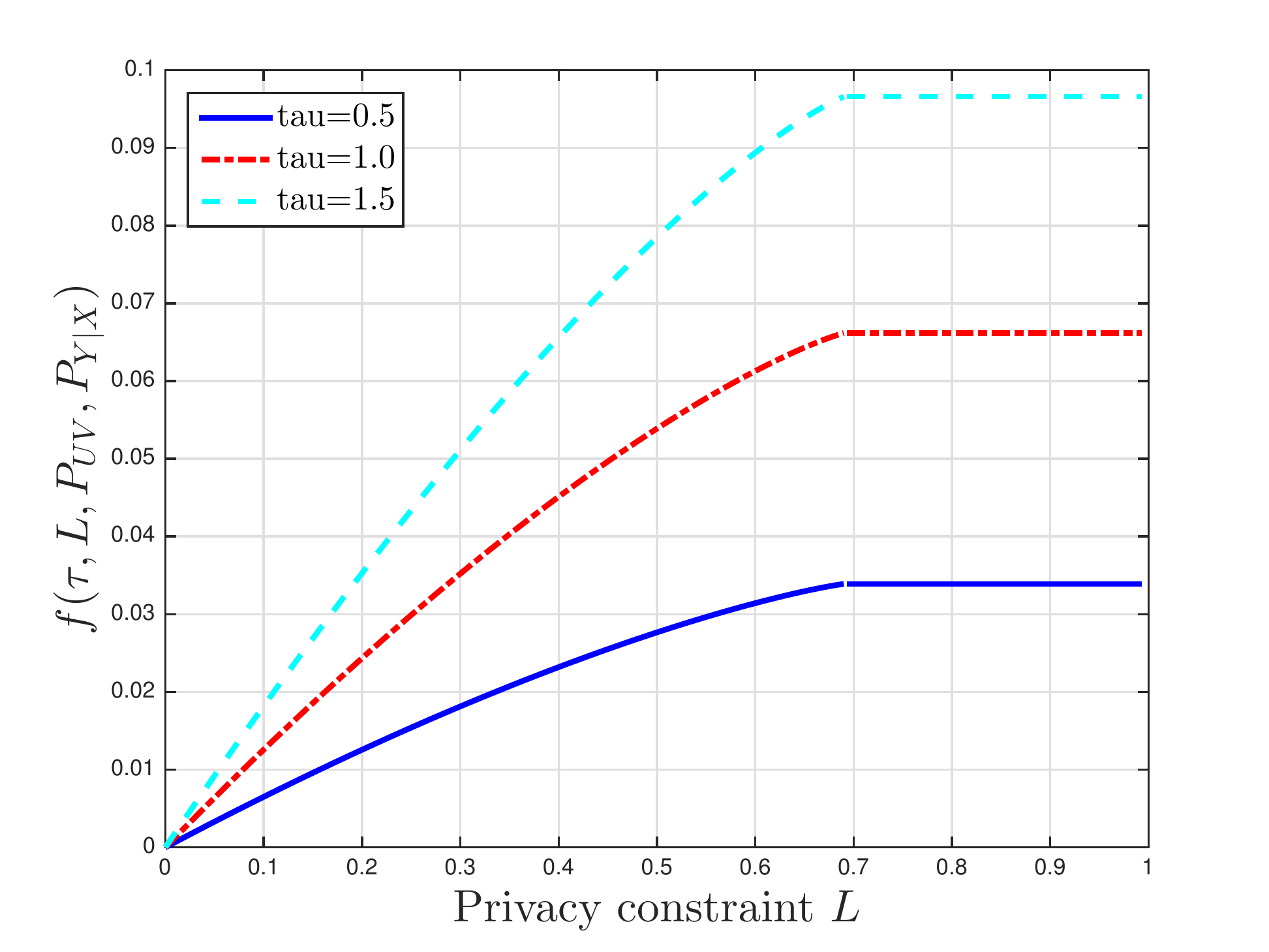}
\caption{Illustration of the privacy-utility tradeoff for a correlated binary source. Here we consider the uniformly distribute binary source $U$ and the side information $V$ is generated by passing $U$ over a binary symmetric channel (BSC) with crossover probability $q$. The noisy channel between the transmitter and the detector is a BSC with crossover probability $0.2$. We optimize the privacy-utility tradeoff over all binary memoryless privacy mechanisms $P_{Z|U}$, which is simple another BSC with a certain crossover probability.}
\label{ill:put}
\end{figure}

Let $\calU=\calV=\calZ=\{1,2\}$. Let $P_U$ be the uniform distribution over $\calU$ and let the transition probability $P_{V|U}$ be 
\begin{align}
P_{V|U}(v|u)=q1_{\{v=u\}}+(1-q)1_{\{v\neq u\}},
\end{align}
for some $q\in[0,1]$. Let the channel $P_{Y|X}$ be a binary symmetric channel with crossover probability $0.2$ and let the privacy mechanism $P_{Z|U}$ be a binary symmetric channel with parameter $p$, which is later optimized over all choices of $p$ to obtain the best privacy mechanism. Using \cite[Proposition 1]{gilani2019distributed}, we can obtain the exact formula of $f(\tau,L,P_{UV},P_{Z|U},P_{Y|X})$. In Figure \ref{ill:put}, we plot the privacy-utility tradeoff for $q=0.8$ and various values of $\tau$. Note that $f(\tau,L,P_{UV},P_{Z|U},P_{Y|X})$ attains the maximal value for any $L\geq H(P_U)=\log 2$ and $f(\tau,L,P_{UV},P_{Z|U},P_{Y|X})=0$ if $L=0$. For any non-degenerate values of $L\in(0,H(P_U))$, we observe a privacy-utility tradeoff.

\subsection{Approximation to the PUT under the High Privacy Limit}
\label{sec:eit}

Note that the exact PUT presented in \eqref{strong:converse} is a non-convex optimization problem, which can not be solved efficiently. In this subsection, under the high privacy limit, i.e., when $I(P_U,P_{Z|U})$ tends to \emph{zero}, we derive an easily computable approximation to the PUT using Euclidean information theory~\cite{borade2008}. Furthermore, as argued in \cite{liao2017hypothesis}, the PUT under the high privacy limit is desirable as we always seek privacy mechanism as strong as possible.

Recall that both $\calU$ and $\calZ$ are finite alphabets. Without loss of generality, in this subsection, we let $\calU=[|\calU|]=\{1,\ldots,|\calU|\}$ and let $\calZ=[|\calZ|]$. Furthermore, we let $\calW:=[|\calW|]=[|\calZ|+1]$. Under the perfect privacy, i.e., $L=0$, we conclude that the privacy mechanism is $P_{Z|U=u}=Q_Z$ for each $u\in\calU$ where $Q_Z\in\calP(\calZ)$ is arbitrary. 
Furthermore, given any $P_{W|Z}$, let $\barQ_W$ be induced by $Q_Z$ and $P_{W|Z}$, i.e.,
\begin{align}
\barQ_W(w)=\sum_z Q_Z(z)P_{W|Z}(w|z)\label{def:QW}.
\end{align}
Given any two finite alphabets $\calA,\calB$ and any distribution $P_A\in\calP(\calA)$, let $\calJ(\calA,\calB,P_A)$ be the collection of all $|\calA|\times|\calB|$ matrices $\bJ=\{J(a,b)\}_{a\in\calA,b\in\calB}$ such that
\begin{align}
|J(a,b)|&\leq 1,~\forall~(a,b)\in\calA\times\calB,\\
\sum_{b\in\calB} J(a,b)&=0,~\forall~a\in\calA,\\
\sum_{a\in\calA}P_A(a)J(a,b)&=0,~\forall~b\in\calB.
\end{align}

Let $\bJ\in\calJ(\calU,\calZ,P_U)$ be an arbitrary. For any $(v,w)$, define
\begin{align}
\nn&h(\bJ,\rho)\\*
&:=\frac{\rho^2}{2}\sum_{v,w}\frac{P_V(v)}{\barQ_W(w)}\Big(\sum_{u,z}P_{U|V}(u|v)P_{W|Z}(w|z)J(u,z)\Big)^2\label{def:hvw},
\end{align}
where we use $J(u,z)$ to denote the $u$-th element of $z$-th row of the matrix $\bJ$.

\begin{figure}[tb]
\centering
\includegraphics[width=.9\columnwidth]{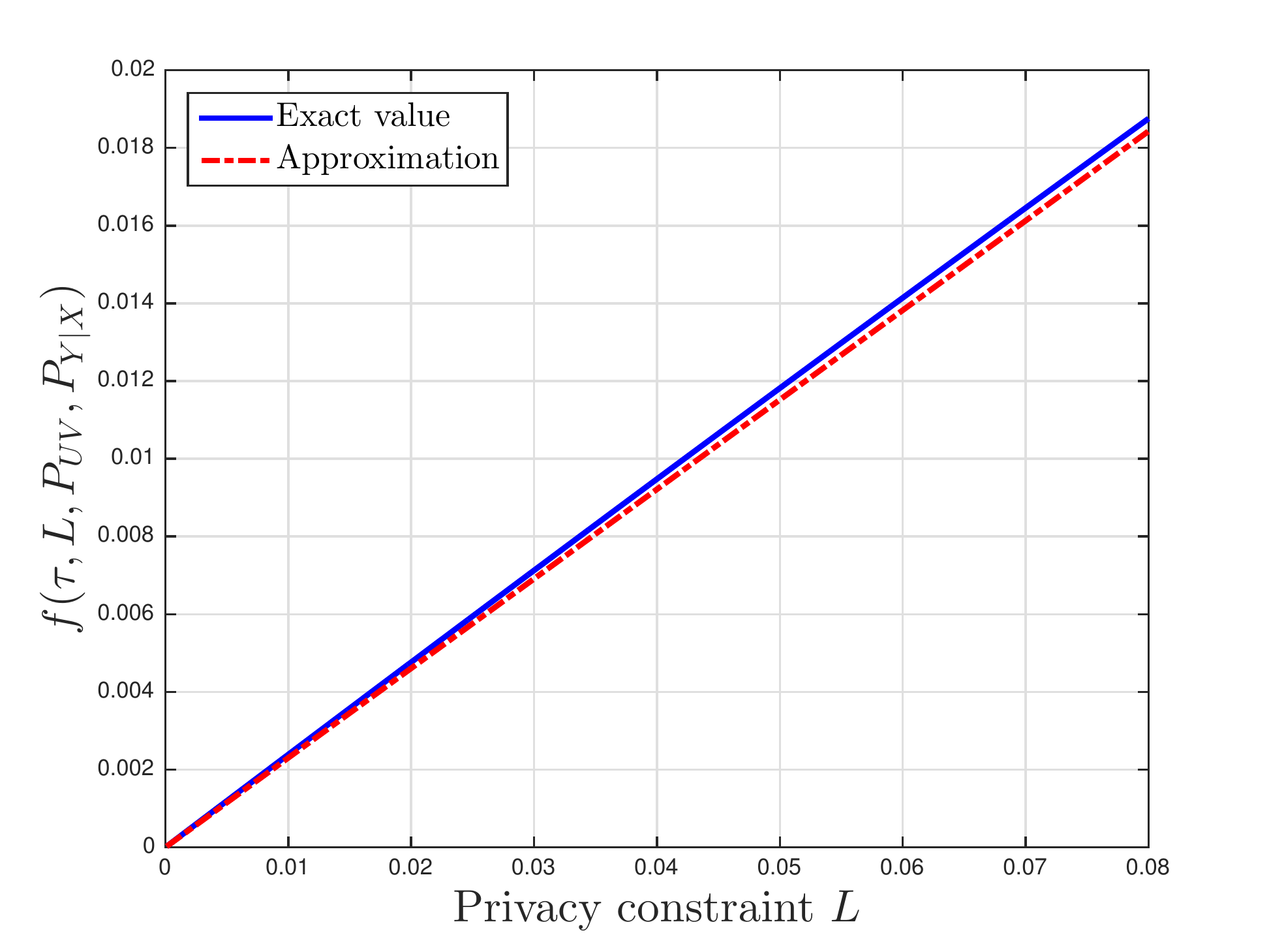}
\caption{Comparison of exact and approximate values for the privacy-utility tradeoff of a binary example. Here we set $L=\frac{\rho^2}{2}$ and $\tau=2$. Note that in this case, the PUT $f\left(\tau,\frac{\rho^2}{2},P_{UV},P_{Y|X}\right)$ increases linearly in $L$.}
\label{illus:approximation2}
\end{figure}

Under the high privacy limit, $L$ can be chosen as $\frac{1}{2}\rho^2$ for an arbitrary small $\rho\in(0,1)$. Using Euclidean information theory~\cite{borade2008,huang2015euclidean}, we have that when $\rho$ is small,
\begin{align}
\nn&f\left(\tau,\frac{\rho^2}{2},P_{UV},P_{Y|X}\right)\\*
&\approx\max_{\substack{Q_Z,P_{W|Z},\bJ\in\calJ(\calU,\calZ,P_U):
\\I(Q_Z,P_{W|Z})\leq \tau C(P_{Y|X})\\\sum_{u,z:Q_Z(z)>0}\frac{P_U(u)(J(u,z))^2}{Q_Z(z)}\leq 1
}}
h(\bJ,\rho)\label{needjust}.
\end{align}
In Figure \ref{illus:approximation2}, the approximation value in \eqref{needjust} is plotted and compared with the exact value for the binary example considered in Section \ref{sec:nume}. We observe that the Euclidean approximation in \eqref{needjust} is quite tight when the privacy constraint $L$ is small.

We then consider the case where the channel $P_{Y|X}$ is extremely noisy so that $ C(P_{Y|X})$ is arbitrarily small. Let $Q_W\in\calP(\calW)$ be an arbitrary distribution, let $\Theta$ be an arbitrary $|\calW|\times|\calZ|$ matrix and define
\begin{align}
\nn& l(\bJ,\Theta,\rho,Q_Z,Q_W)\\*
&:=\frac{\rho^4}{2}\sum_{v,w}\frac{P_V(v)}{Q_W(w)}\Big(\sum_{u,z}P_{U|V}(u|v)J(u,z)\Theta(z,w)\Big)^2,
\end{align}
where we use $\Theta(z,w)$ to denote the $z$-th element of $w$-th row of the matrix $\Theta$.

If we further assume that the channel $P_{Y|X}$ is extremely noisy such that $\tau C(P_{Y|X})=\frac{\rho^2}{2}$, then 
\begin{align}
\nn&f\left(\tau,\frac{\rho^2}{2},P_{UV},P_{Y|X}\right)\\*
&\approx \max_{\substack{Q_Z,Q_W,
\Theta\in\calJ(\calZ,\calW,Q_Z),\bJ\in\calJ(\calU,\calZ,P_U)\\
\\\sum_{z,w:Q_W(w)>0}\frac{Q_Z(z)(\Theta(z,w))^2}{Q_W(w)}\leq 1\\
\sum_{u,z:Q_Z(z)>0}\frac{P_U(u)(J(u,z))^2}{Q_Z(z)}\leq 1
}}l(\bJ,\Theta,\rho,Q_Z,Q_W)\label{needjust2}.
\end{align}
The proofs of \eqref{needjust} and \eqref{needjust2} are provided in Appendix \ref{proof:needjust}.

In Figure \ref{illus:approximation}, the approximation value given in \eqref{needjust2} is plotted and compared with the exact value for the binary example considered in Section \ref{sec:nume}. We observe that the Euclidean approximation in \eqref{needjust2} is very tight when the privacy constraint $L$ is small and the channel is extremely noisy.
\begin{figure}[tb]
\centering
\includegraphics[width=.9\columnwidth]{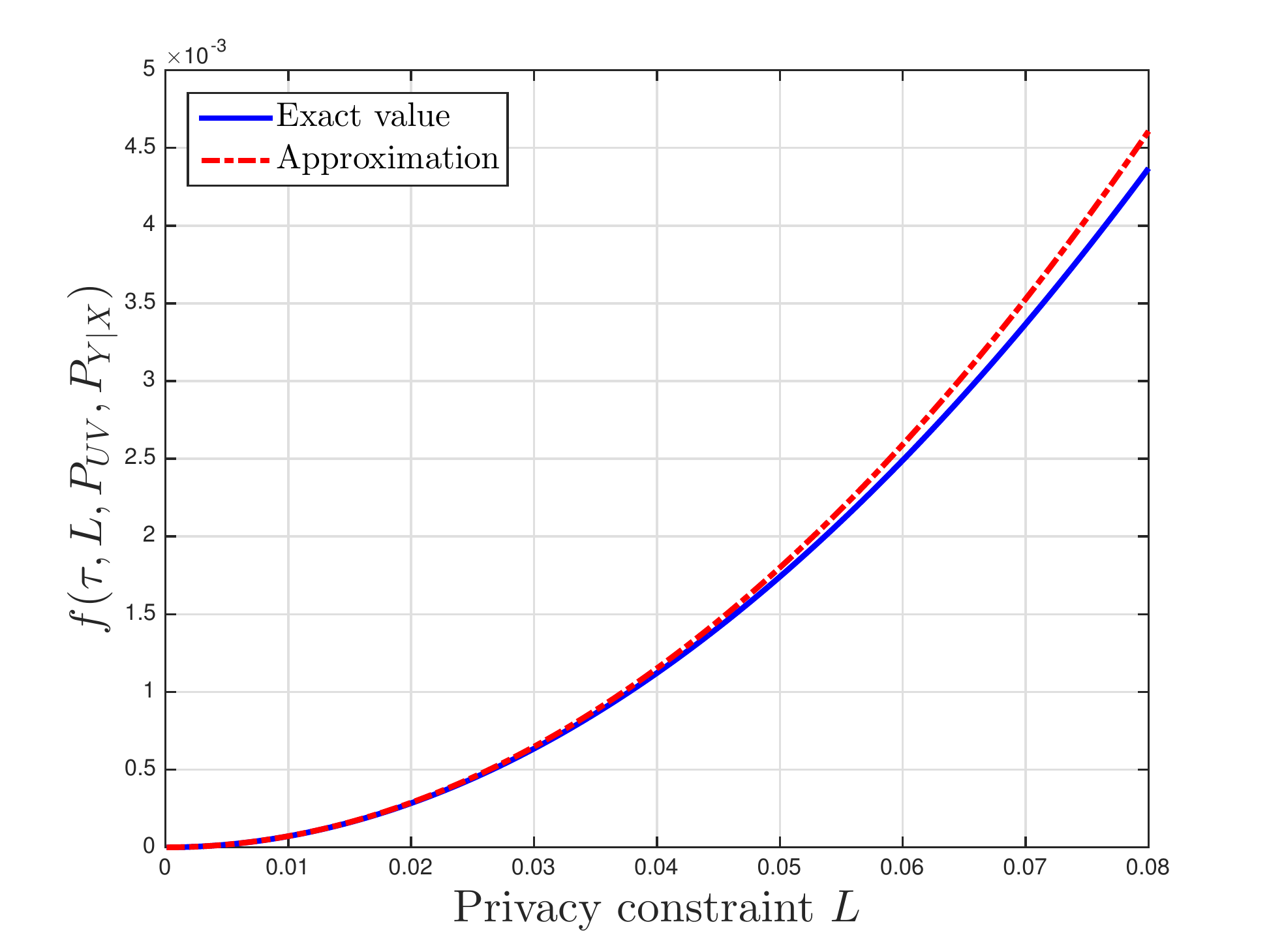}
\caption{Comparison of exact and approximate values for the privacy-utility tradeoff of a binary example. Here we set $\tau C(P_{Y|X})=L=\frac{\rho^2}{2}$ so that $\rho^4=4L^2$. Note that in this case, the PUT $f\left(\tau,\frac{\rho^2}{2},P_{UV},P_{Y|X}\right)$ increases quadratically in $L$.}
\label{illus:approximation}
\end{figure}

\section{Proof of Theorem \ref{mainresult}}
\label{sec:proof}

\subsection{Alternative Characterization of the Optimal PUT}
We first provide an alternative characterization of the optimal privacy-utility tradeoff $f(\tau,L,P_{UV},P_{Z|U},P_{Y|X})$ in \eqref{def:putf} using the supporting hyperplanes~\cite{oohama2016wynerziv,oohama2015wak}. This result is critical to our converse proof.

Recall that $P_{UV}$ is the generating distribution of $(U^k,V^k)$ under hypothesis $\rmH_1$ and $P_{Y|X}$ denotes the memoryless channel between the transmitter and the detector. For any memoryless privacy mechanism $P_{Z|U}$, let $P_U$, $P_Z$, $P_{U|Z}$ and $P_{V|U}$ be distributions induced by $P_{UV}$ and $P_{Z|U}$. Furthermore, recall that $\calQ$ denotes the set of all joint distributions defined on the alphabet $\calU\times\calV\times\calZ\times\calW$ and that $\calC$ denotes the set of all joint distributions defined on the alphabet $\calX\times\calY$. Given any $(Q_{UVZW},Q_{XY})\in\calQ\times\calC$, for any $(\lambda_1,\lambda_2,\gamma)\in\bbR_+^3$, let
\begin{align}
\nn&\Delta_{\gamma}^{\tau,L}(Q_{UVZW},Q_{XY},P_{UV},P_{Z|U},P_{Y|X})\\*
\nn&:=\gamma D(Q_Z\|P_Z)+\gamma D(Q_{UV|ZW}\|P_{U|Z}P_{V|U}|Q_{ZW})\\*
&\qquad+\tau\gamma D(Q_{Y|X}\|P_{Y|X}|Q_X),\label{def:Deltabg}\\
\nn&R_{\lambda_1,\lambda_2,\gamma}^{\tau,L}(Q_{UVZW},Q_{XY},P_{UV},P_{Z|U},P_{Y|X})\\*
&:=R_{\lambda_1,\lambda_2}^{\tau,L}(Q_{UVZW},Q_{XY})\nn\\*
&\quad\quad -\Delta_{\gamma}^{\tau,L}(Q_{UVZW},Q_{XY},P_{UV},P_{Z|U},P_{Y|X})
\label{def:rbgQ},
\end{align}
where $R_{\lambda_1,\lambda_2}^{\tau,L}(Q_{UVZW},Q_{XY})$ was defined in \eqref{def:RbQ}.

Finally, let
\begin{align}
\nn&g_{\lambda_1,\lambda_2,\gamma}^{\tau,L}(P_{UV},P_{Z|U},P_{Y|X})\\*
&:=\sup_{\substack{Q_{UVZW}\in\calQ\\Q_{XY}\in\calC}}R_{\lambda_1,\lambda_2,\gamma}^{\tau,L}(Q_{UVZW},Q_{XY},P_{UV},P_{Z|U},P_{Y|X})\label{def:rbg}.
\end{align}

Recall the definitions of $\zeta(\lambda_1,\lambda_2,\gamma,\tau)$ in \eqref{def:cbgt}, $f(\tau,L,P_{UV},P_{Z|U},P_{Y|X})$ in \eqref{def:putf} and $g_{\lambda_1,\lambda_2}^{\tau,L}(\cdot)$ in \eqref{def:rb}. We have the following lemma.
\begin{lemma}
\label{alt:expression}
The following claims hold:
\begin{enumerate}
\item $g_{\lambda_1,\lambda_2}^{\tau,L}(\cdot)$ is related with $f_{\lambda_1,\lambda_2}^{\tau,L}(\cdot)$ as follows:
\begin{align}
f(\tau,L,\cdot)=\min_{(\lambda_1,\lambda_2)\in\bbR_+^2}g_{\lambda_1,\lambda_2}^{\tau,L}(\cdot)\label{eqn:alt},
\end{align}
\item $g_{\lambda_1,\lambda_2,\gamma}^{\tau,L}(\cdot)$ is related with $g_{\lambda_1,\lambda_2}^{\tau,L}(\cdot)$ as follows:
\begin{align}
g_{\lambda_1,\lambda_2,\gamma}^{\tau,L}(\cdot)
&\geq 
g_{\lambda_1,\lambda_2}^{\tau,L}(\cdot),\\*
g_{\lambda_1,\lambda_2,\gamma}^{\tau,L}(\cdot)&\leq g_{\lambda_1,\lambda_2}^{\tau,L}(\cdot)+\zeta(\lambda_1,\lambda_2,\gamma,\tau),
\end{align}
\end{enumerate}
where $\cdot$ denotes the triple of (conditional) distributions $P_{UV},P_{Z|U},P_{Y|X}$.
\end{lemma}
The proof of Lemma \ref{alt:expression} uses the Lagrange multiplier method in convex optimization~\cite{boyd2004convex} and is provided in Appendix \ref{proof:alt:expression}.

\subsection{Equivalent Expressions for Error Probabilities}
Fix any $k\in\bbN$ and consider any $n\leq \tau k$. Given a memoryless privacy mechanism $P_{Z|U}^k$ and a communication protocol with a potentially stochastic encoder $f^{n,k}$ and a decoder $g^{n,k}$, define the following joint distributions:
\begin{align}
\nn&P_{U^kV^kZ^kX^nY^n}(u^k,v^k,z^k,x^n,y^n)\\*
&=P_{UV}^k(u^k,v^k)P_{Z|U}^k(z^k|u^k)P_{f^{n,k}}(x^n|z^k) P_{Y|X}^n(y^n|x^n)\label{def:P},\\
\nn&Q_{U^kV^kZ^kX^nY^n}(u^k,v^k,z^k,x^n,y^n)\\*
&=P_U^k(u^k)P_V^k(v^k)P_{Z|U}^k(z^k|u^k)P_{f^{n,k}}(x^n|z^k)P_{Y|X}^n(y^n|x^n)\label{def:Q},
\end{align}
where $P_{f^{n,k}}(x^n|z^k)$ denotes the probability that the output of the encoder is $x^n$ when the input is $z^k$.

Define the acceptance region
\begin{align}
\calA&:=\{(y^n,v^k):~g^{n,k}(y^n,v^k)=\rmH_1\}\label{def:aregion}.
\end{align}
Furthermore, let $P_{Z^k}$, $P_{Y^n}$, $P_{U^kZ^k}$, $P_{Y^nV^k}$ and $P_{Y^nV^k|U^kZ^kX^n}$ be induced by the joint distribution $P_{U^kV^kZ^kX^nY^n}$ and let $Q_{Y^nV^k}$ be induced by $Q_{U^kV^kZ^kX^nY^n}$. Note that the marginal distribution of $(U^k,Z^k)$ is $P_{UZ}^k$ and the marginal distribution of $V^k$ is $P_V^k$ under both distributions $P_{U^kV^kZ^kX^nY^n}$ and $Q_{U^kV^kZ^kX^nY^n}$. The marginal distribution of $Y^n$ is the same under both joint distributions and denoted as $P_{Y^n}$, i.e.,
\begin{align}
P_{Y^n}(y^n)
\nn&:=\sum_{u^k,z^k,x^n}P_U^k(u^k)P_{Z|U}^k(z^k|u^k)\\
&\qquad \qquad \times P_{f^{n,k}}(x^n|z^k)P_{Y|X}^n(y^n|x^n).
\end{align}

Then the type-I and type-II error probabilities are equivalently expressed as follows:
\begin{align}
\beta_1(f^{n,k},g^{n,k})&=P_{Y^nV^k}(\calA^\rmc)\label{type1:eqn},\\
\beta_2(f^{n,k},g^{n,k})&=Q_{Y^nV^k}(\calA)\label{type2:eqn}.
\end{align}

\subsection{Construct the Truncated Distribution}
We consider any memoryless privacy mechanism $P_{Z|U}$ and any communication protocol $(f^{n,k},g^{n,k})$ such that i) the privacy constraint is satisfied with parameter $L$ and ii) the type-I error probability is upper bounded by $\varepsilon\in(0,1)$, i.e.,
\begin{align}
I(P_U,P_{Z|U})&\leq L\label{privacy:constraint},\\
\beta_1(f^{n,k},g^{n,k})&\leq \varepsilon\label{type1:constraint}.
\end{align}

Define a set concerning the detection probability at the decoder
\begin{align}
\calB_1
\nn&:=\bigg\{(u^k,z^k,x^n):\\*
&\quad P_{Y^nV^k|U^kZ^kX^n}(\calA|u^k,z^k,x^n)\geq \frac{1-\varepsilon}{4}\bigg\}\label{def:calB}.
\end{align}
Then we have,
\begin{align}
\nn&1-\varepsilon\\*
&\leq P_{Y^nV^k}(\calA)\label{usetwo}\\
&=\sum_{\substack{u^k,v^k,z^k,x^n,y^n:\\(v^k,y^n)\in\calA}}P_{U^kV^kZ^kX^nY^n}(u^k,v^k,z^k,x^n,y^n)\\
\nn&=\sum_{u^k,z^k,x^n}P_{UZ}^k(u^k,z^k)P_{f^{n,k}}(x^n|z^k)\\*
&\qquad\qquad\times P_{Y^nV^k|U^kZ^kX^n}(\calA|u^k,z^k,f^{n,k}(z^k))\\
\nn&=\sum_{(u^k,z^k,x^n)\in\calB_1}P_{UZ}^k(u^k,z^k)P_{f^{n,k}}(x^n|z^k)\\*
\nn&\qquad\qquad\qquad\qquad\times P_{Y^nV^k|U^kZ^kX^n}(\calA|u^k,z^k,x^n)\\*
\nn&+\sum_{(u^k,z^k,x^n)\notin\calB_1}P_{UZ}^k(u^k,z^k)P_{f^{n,k}}(x^n|z^k)\\*
&\qquad\qquad\qquad\qquad\times P_{Y^nV^k|U^kZ^kX^n}(\calA|u^k,z^k,x^n)\\
&\leq P_{U^kZ^kX^n}(\calB_1)+\frac{1-\varepsilon}{4}\label{usedefB1ha},
\end{align}
where \eqref{usetwo} follows from the equivalent expression of the type-I error probability in \eqref{type1:eqn} and the constraint on the type-I error probability in \eqref{type1:constraint}, and \eqref{usedefB1ha} follows from the definition of $\calB_1$ in \eqref{def:calB}.
Thus,
\begin{align}
P_{U^kZ^kX^n}(\calB_1)\geq \frac{3(1-\varepsilon)}{4}\label{result1}.
\end{align}

Recall the definitions of $\imath(u;z|P_{UZ})$ in \eqref{def:mi} and $L(P_{Z|U},\varepsilon)$ in \eqref{def:L}. Define another set concerning the privacy constraint
\begin{align}
\calB_2
\nn:=\bigg\{(u^k,z^k,x^n):\sum_{i\in[k]}&\imath(u_i;z_i|P_{Z|U})\leq k I(P_U,P_{Z|U})\\*
&+\sqrt{k}L(P_{Z|U},(1-\varepsilon)/4)\bigg\}\label{def:calB2}.
\end{align}
Applying the Berry-Esseen theorem~\cite{berry1941accuracy,esseen1942liapounoff}, we have
\begin{align}
P_{U^kZ^kX^n}(\calB_2^\rmc)\leq \frac{1-\varepsilon}{4}\label{result2}.
\end{align}

Define the intersection of two sets as
\begin{align}
\calB&:=\calB_1\cap\calB_2\label{def:calBint}.
\end{align}
The results in \eqref{result1} and \eqref{result2} imply that
\begin{align}
P_{U^kZ^kX^n}(\calB)
&=P_{U^kZ^kX^n}(\calB_1\cap\calB_2)\\
&=1-P_{U^kZ^kX^n}(\calB_1^c\cup\calB_2^c)\\
&\geq 1- P_{U^kZ^kX^n}(\calB_1^c)-P_{U^kZ^kX^n}(\calB_2^c)\\
&\geq 1-\left(1-\frac{3(1-\varepsilon)}{4}\right)-\frac{1-\varepsilon}{4}\\
&\geq \frac{1-\varepsilon}{2}.
\end{align}

Consider random variables $(\tilU^k,\tilZ^k,\tilV^k,\tilX^n,\tilY^n)$ with joint distribution $P_{\tilU^k\tilZ^k\tilV^k\tilX^n\tilY^n}$ such that 
\begin{align}
\nn&P_{\tilU^k\tilZ^k\tilV^k\tilX^n\tilY^n}(u^k,v^k,z^k,x^n,y^n)\\*
\nn&=\frac{P_U^k(u^k)P_{Z|U}^k(z^k|u^k)P_{f^{n,k}}(x^n|z^k)1\{(u^k,z^k,x^n)\in\calB\}}{P_{U^kZ^kX^n}(\calB)}\\*
&\times\frac{P_{Y^nV^k|U^kZ^kX^n}(y^n,v^k|u^k,z^k,x^n)1\{(y^n,v^k)\in\calA\}}{P_{Y^nV^k|U^kZ^kX^n}(\calA|u^k,z^k,x^n)}\label{def:Ptil}.
\end{align}

Note that the joint distribution in \eqref{def:Ptil} is a truncated distribution of the original one $P_{U^kZ^kV^kX^nY^n}$, by considering only $(u^k,z^k,x^n)\in\calB$ and $(y^n,v^k)\in\calA$. The truncated distribution in \eqref{def:Ptil} allows us to apply change-of-measure and then use the strong converse technique introduced in \cite{tyagi2020strong}. As we shall show shortly below, under the truncated distribution, the type-I error probability is zero and the truncated distribution in \eqref{def:Ptil} is close to the original distribution in terms of KL divergence.

Let $P_{\tilY^n}$, $P_{\tilV^k}$ and $P_{\tilY^n\tilV^k}$ be induced by $P_{\tilZ^k\tilV^k\tilX^n\tilY^n}$. From \eqref{def:Ptil}, we have
\begin{align}
P_{\tilY^n\tilV^k}(\calA)=1\label{result22}.
\end{align}
Note that the constructed distribution $P_{\tilU^k\tilZ^k\tilV^k\tilX^n\tilY^n}$ is in fact close to the distribution $P_{Z^kV^kX^nY^n}$ in terms of KL divergence, i.e.,
\begin{align}
\nn&D(P_{\tilU^k\tilZ^k\tilV^k\tilX^n\tilY^n}\|P_{U^kZ^kV^kX^nY^n})\\*
\nn&=\sum_{u^k,v^k,z^k,x^n,y^n}P_{\tilU^k\tilZ^k\tilV^k\tilX^n\tilY^n}(u^k,v^k,z^k,x^n,y^n)\\*
&\qquad\times\log\frac{P_{\tilU^k\tilZ^k\tilV^k\tilX^n\tilY^n}(u^k,v^k,z^k,x^n,y^n)}{P_{U^kZ^kV^kX^nY^n}(u^k,v^k,z^k,x^n,y^n)}\\
\nn&=\sum_{u^k,v^k,z^k,x^n,y^n}P_{\tilU^k\tilZ^k\tilV^k\tilX^n\tilY^n}(u^k,v^k,z^k,x^n,y^n)\\*
&\qquad\times\log\frac{1}{P_{U^kZ^kX^n}(\calB)P_{Y^nV^k|U^kZ^kX^n}(\calA|u^k,z^k,x^n)}\\
&\leq -2\log(1-\varepsilon)+3\log 2\label{useresults},
\end{align}
where \eqref{useresults} follows from the definition of $\calB$ in \eqref{def:calB} and the result in \eqref{result1}.

\subsection{Multiletter Bound for PUT}
Let $P_{\tilU^k}$, $P_{\tilZ^k}$ and $P_{\tilU^k\tilZ^k}$ be induced by $P_{\tilU^k\tilZ^k\tilV^k\tilX^n\tilY^n}$. It follows that
\begin{align}
\nn&I(\tilU^k;\tilZ^k)\\*
&=\mathbb{E}_{P_{\tilU^k\tilZ^k}}\left[\log\frac{P_{\tilU^k\tilZ^k}(U^k,Z^k)}{P_{\tilU^k}(U^k)P_{\tilZ^k}(Z^k)}\right]\\
\nn&=\mathbb{E}_{P_{\tilU^k\tilZ^k}}\bigg[\log\bigg(\frac{P_{\tilU^k\tilZ^k}(U^k,Z^k)}{P_{UZ}^k(U^k,Z^k)}\frac{P_U^k(U^k)P_Z^k(Z^k)}{P_{\tilU^k}(U^k)P_{\tilZ^k}(Z^k)}\\*
&\qquad\qquad\qquad\times\frac{P_{UZ}^k(U^k,Z^k)}{P_U^k(U^k)P_Z^k(Z^k)}\bigg)\bigg]\\
\nn&=D(P_{\tilU^k\tilZ^k}\|P_{UZ}^k)-D(P_{\tilU^k}\|P_U^k)-D(P_{\tilZ^k}\|P_Z^k)\\*
&\qquad+\mathbb{E}_{P_{\tilU^k\tilZ^k}}\Big[\sum_{i\in[k]}\imath(U_i;Z_i|P_{Z|U})\Big]\label{usemi}\\
\nn&\leq D(P_{\tilU^k\tilZ^k}\|P_{UZ}^k)+k I(P_U,P_{Z|U})\\*
&\qquad+\sqrt{k}L(P_{Z|U},(1-\varepsilon)/4)\label{usecalB2}\\
\nn&\leq D(P_{\tilU^k\tilZ^k\tilX^k}\|P_{UZX}^k)+k I(P_U,P_{Z|U})\\*
&\qquad+\sqrt{k}L(P_{Z|U},(1-\varepsilon)/4)\label{dpkl}\\
&\leq \log\frac{2}{1-\varepsilon}+k I(P_U,P_{Z|U})+\sqrt{k}L(P_{Z|U},(1-\varepsilon)/4)\label{usepartialresults}\\
&\leq kL+\log\frac{2}{1-\varepsilon}+\sqrt{k}L(P_{Z|U},(1-\varepsilon)/4)\label{useprivacycons},
\end{align}
where \eqref{usemi} follows from the definition of $\imath(u;z|P_{Z|U})$ in \eqref{def:mi}, \eqref{usecalB2} follows from the definitions of $\calB_2$ in \eqref{def:calB2} and $\calB$ in \eqref{def:calB}, \eqref{dpkl} follows from the data processing inequality for the KL divergence, \eqref{usepartialresults} follows similarly to \eqref{useresults} and \eqref{useprivacycons} follows from the privacy constraint in \eqref{privacy:constraint}.

We then derive an upper bound on the type-II error exponent. Using \eqref{type2:eqn}, we have
\begin{align}
\nn&-\log\beta_2(f^{n,k},g^{n,k})\\*
&=-\log Q_{Y^nV^k}(\calA)\\
&=P_{\tilY^n\tilV^k}(\calA)\log\frac{P_{\tilY^n\tilV^k}(\calA)}{Q_{Y^nV^k}(\calA)}\label{usep=1}\\
&\leq \sum_{(y^n,v^k)\in\calA}P_{\tilY^n\tilV^k}(y^n,v^k)\log\frac{P_{\tilY^n\tilV^k}(y^n,v^k)}{Q_{Y^nV^k}(y^n,v^k)}\label{uselogsum}\\
&=\sum_{(y^n,v^k)}P_{\tilY^n\tilV^k}(y^n,v^k)\log\frac{P_{\tilY^n\tilV^k}(y^n,v^k)}{Q_{Y^nV^k}(y^n,v^k)}\label{usep=1again}\\
&=D(P_{\tilY^n\tilV^k}\|Q_{Y^nV^k})\\
&=D(P_{\tilY^n\tilV^k}\|P_{Y^n}P_V^k)\label{usedefQ}\\
\nn&=D(P_{\tilY^n\tilV^k}\|P_{\tilY^n}P_{\tilV^k})\\*
&\qquad+\sum_{y^n,v^k}P_{\tilY^n\tilV^k}(y^n,v^k)\log\frac{P_{\tilY^n}(y^n)P_{\tilV^k}(v^k)}{P_{Y^n}(y^n)P_V^k(v^k)}\\
&\leq D(P_{\tilY^n\tilV^k}\|P_{\tilY^n}P_{\tilV^k})+\log\frac{64}{(1-\varepsilon)^4}\label{toexplain}\\
&=I(\tilY^n;\tilV^k)-4\log(1-\varepsilon)+6\log 2\label{lb1},
\end{align}
where \eqref{usep=1} and \eqref{usep=1again} follow from the fact that $P_{\tilY^n\tilV^k}(\calA)=1$ in \eqref{result22}, \eqref{uselogsum} follows from the log-sum inequality, \eqref{usedefQ} follows from the definition of $Q_{Y^nV^k}$ (cf. \eqref{def:Q}), and \eqref{toexplain} follows since from \eqref{def:calB}, \eqref{result1} and \eqref{def:Ptil}, 
\begin{align}
\nn&P_{\tilY^n}(y^n)\\*
&=\sum_{u^k,v^k,z^k,x^n}P_{\tilU^k\tilZ^k\tilV^k\tilX^n\tilY^n}(u^k,v^k,z^k,x^n,y^n)\\
\nn&\leq\sum_{u^k,v^k,z^k,x^n}\frac{P_U^k(u^k)P_{Z|U}^k(z^k|u^k)P_{f^{n,k}}(x^n|z^k)}{P_{U^kZ^kX^n}(\calB)}\\*
&\qquad\times\frac{P_{Y^nV^k|U^kZ^kX^n}(y^n,v^k|u^k,z^k,x^n)}{P_{Y^nV^k|U^kZ^kX^n}(\calA|z^k,x^n)}\\
&\leq\frac{P_{Y^n}(y^n)}{P_{U^kZ^kX^n}(\calB)P_{Y^nV^k|U^kZ^kX^n}(\calA|z^k,x^n)}\\
&\leq \frac{8P_{Y^n}(y^n)}{(1-\varepsilon)^2},
\end{align}
and similarly $P_{\tilV^k}(v^k)\leq \frac{8P_V^k(v^k)}{(1-\varepsilon)^2}$.

Recall the joint distribution of $(\tilU^k,\tilZ^k,\tilV^k,\tilX^n,\tilY^n)$ in \eqref{def:Ptil}. We have
\begin{align}
\nn&I(\tilZ^k,\tilV^k;\tilY^n)-I(\tilX^n;\tilY^n)\\*
&\leq I(\tilZ^k,\tilV^k,\tilX^n;\tilY^n)-I(\tilX^n;\tilY^n)\label{usetilPf}\\
&=I(\tilZ^k,\tilV^k;\tilY^n|\tilX^n)\\*
&=D(P_{\tilY^n|\tilZ^k\tilV^k\tilX^n}\|P_{\tilY^n|\tilX^n}|P_{\tilZ^k\tilV^k\tilX^n})\\
\nn&=D(P_{\tilY^n|\tilZ^k\tilV^k\tilX^n}\|P_{Y^n|Z^kV^kX^n}|P_{\tilZ^k\tilV^k\tilX^n})\\*
&\qquad-D(P_{\tilY^n|\tilX^n}\|P_{Y|X}^n|P_{\tilX^n})\label{useP}\\
&\leq D(P_{\tilY^n|\tilZ^k\tilV^k\tilX^n}\|P_{Y^n|Z^kV^kX^n}|P_{\tilZ^k\tilV^k\tilX^n})\\
&\leq D(P_{\tilU^k\tilZ^k\tilV^k\tilX^n\tilY^n}\|P_{U^kZ^kV^kX^nY^n})\\
&\leq -2\log(1-\varepsilon)+3\log 2\label{usedlsmall},
\end{align}
where \eqref{useP} follows from the Markov chain $Y^n-X^n-(Z^k,V^k)$ under the joint distribution $P_{Z^kV^kX^nY^n}$ (cf. \eqref{def:P}) and \eqref{usedlsmall} follows from \eqref{useresults}.

Combining \eqref{useresults}, \eqref{useprivacycons}, \eqref{lb1} and \eqref{usedlsmall}, for any $(\lambda_1,\lambda_2)\in\bbR_+^2$, we have
\begin{align}
\nn&-\log\beta_2(f^{n,k},g^{n,k})\\*
\nn&\leq I(\tilY^n;\tilV^k)-\lambda_1(I(\tilZ^k,\tilV^k;\tilY^n)-I(\tilX^n;\tilY^n))\\*
\nn&\qquad-\lambda_2(I(\tilU^k;\tilZ^k)-kL)
-(2\lambda_1+\lambda_2+2\gamma)\log(1-\varepsilon)\\*
\nn&\qquad+(3\lambda_1+\lambda_2+3\gamma)\log 2\\*
\nn&\qquad-\gamma D(P_{\tilU^k\tilZ^k\tilV^k\tilX^n\tilY^n}\|P_{U^kZ^kV^kX^nY^n})\\*
&\qquad+\lambda_2\sqrt{k}L(P_{Z|U},(1-\varepsilon)/4)\label{nletter:ub}.
\end{align}

\subsection{Single Letterize the PUT}
For any $n\in\bbN$, let $J_n$ be the uniform random variable over $[n]$, which is independent of any other random variables. Furthermore, for simplicity, we use $J$ to denote $J_k$. For each $i\in[k]$, let $W_i:=(\tilZ^{i-1},\tilV^{i-1},\tilY^n)$. Using standard single-letterization technique, we have the following lemma.

\begin{lemma}
\label{single-letter:terms}
The following results hold:
\begin{align}
I(\tilU^k;\tilZ^k)&\geq kI(\tilU_J;\tilZ_J)-2\log\frac{2}{1-\varepsilon},\\
I(\tilY^n;\tilV^k)&\leq kI(W_{J},J;\tilV_{J}),\\
I(\tilX^n;\tilY^n)&\leq nI(\tilX_{J_n};\tilY_{J_n},J_n)-2\log(1-\varepsilon)+3\log 2,\\
I(\tilZ^k,\tilV^k;\tilY^n)&\geq kI(\tilZ_{J};W_{J},J)+2\log (1-\varepsilon)-3\log 2,
\end{align}
and
\begin{align}
\nn&D(P_{\tilU^k\tilZ^k\tilV^k\tilX^n\tilY^n}\|P_{U^kZ^kV^kX^nY^n})\\*
\nn&\geq k D(P_{\tilZ_J}\|P_Z)+nD(P_{\tilY_{J_n}|\tilX_{J_n}}\|P_{Y|X}|P_{\tilX_{J_n}})\\*
&\qquad+k D(P_{\tilU_J\tilV_J|\tilZ_J\tilW_J}\|P_{U|Z}P_{V|U}|P_{\tilZ_J\tilW_J}).
\end{align}
\end{lemma}
The proof of Lemma \ref{single-letter:terms} is provided in Appendix \ref{proof:single}.

Define random variables $(U',Z',V',X',Y',W')$ such that $U'=\tilU_J$, $Z'=\tilZ_J$, $V'=V_J$, $X'=X_{J_n}$, $Y'=Y_{J_n}$ and $W'=(W_J,J)$. Using the joint distribution of $(\tilU^k,\tilZ^k,\tilV^k,\tilX^n,\tilY^n)$ in \eqref{def:Ptil} and the definitions of $W_i$, $J$ and $J_n$, we obtain the joint distribution $P_{U'Z'V'W'}$ of random variables $(U',Z',V',W')$ and the joint distribution $P_{X'Y'}$ of random variables $(X',Y')$.  

Combining \eqref{def:rbgQ}, \eqref{nletter:ub} and Lemma \ref{single-letter:terms}, we conclude that given any $(\lambda_1,\lambda_2,\gamma)\in\bbR_+^2$,
for any $(f^{n,k},g^{n,k},P_{Z|U}^k)$ such that $n\leq k\tau$ and $\beta_1(f^{n,k},g^{n,k})\leq \varepsilon$ and $I(P_U,P_{Z|U})\leq L$,
\begin{align}
\nn&-\log\beta_2(f^{n,k},g^{n,k})\\*
\nn&\leq kR_{\lambda_1,\lambda_2,\gamma}^{\tau,L}(P_{U'Z'V'W'},P_{X'Y'},P_{UV},P_{Z|U},P_{Y|X})\\*
\nn&\qquad-(6\lambda_1+3\lambda_2+2\gamma)\log(1-\varepsilon)\\*
\nn&\qquad+(9\lambda_1+3\lambda_2+3\gamma)\log 2\\*
&\qquad+\lambda_2\sqrt{k}L(P_{Z|U},(1-\varepsilon)/4)\\
\nn&\leq kg_{\lambda_1,\lambda_2,\gamma}^{\tau,L}(P_{UV},P_{Z|U},P_{Y|X})\\*
\nn&\qquad-(6\lambda_1+3\lambda_2+2\gamma)\log(1-\varepsilon)\\*
\nn&\qquad+(9\lambda_1+3\lambda_2+3\gamma)\log 2\\*
&\qquad+\lambda_2\sqrt{k}L(P_{Z|U},(1-\varepsilon)/4)\label{useRbg}\\
\nn&\leq kg_{\lambda_1,\lambda_2}^{\tau,L}(P_{UV},P_{Z|U},P_{Y|X})+k\tau \zeta(\lambda_1,\lambda_2,\gamma,\tau)\\*
\nn&\qquad-(6\lambda_1+3\lambda_2+2\gamma)\log(1-\varepsilon)\\*
\nn&\qquad+(9\lambda_1+3\lambda_2+3\gamma)\log 2\\*
&\qquad+\lambda_2\sqrt{k}L(P_{Z|U},(1-\varepsilon)/4)\label{useclaim2},
\end{align}
where \eqref{useRbg} follows from the definition of $g_{\lambda_1,\lambda_2,\gamma}^{\tau,L}(\cdot)$ in \eqref{def:rbg}, \eqref{useclaim2} follows from Claim 2) in Lemma \ref{alt:expression}. 

Let $(\lambda_1^*,\lambda_2^*)$ be an optimizer in \eqref{eqn:alt} such that $g_{\lambda_1,\lambda_2}^{\tau,L}(P_{UV},P_{Z|U},P_{Y|X})=f(\tau,L,P_{UV},P_{Z|U},P_{Y|X})$. Note that both $\lambda_1^*$ and $\lambda_2^*$ are finite. Choosing $\gamma=\sqrt{k}$, using the definition of $\zeta(\lambda_1,\lambda_2,\gamma,\tau)$ in \eqref{def:cbgt} and the result in \eqref{useclaim2}, we have
\begin{align}
\nn&\liminf_{k\to\infty}E^*(k,\tau,L,\varepsilon)\\*
&\leq g_{\lambda_1^*,\lambda_2^*}^{\tau,L}(P_{UV},P_{Z|U},P_{Y|X})\\
&=f(\tau,L,P_{UV},P_{Z|U},P_{Y|X})\\
&\leq \max_{P_{Z|U}}f(\tau,L,P_{UV},P_{Z|U},P_{Y|X}).
\end{align}

\section{Conclusion}
\label{sec:conc}
We derived the privacy-utility tradeoff for a hypothesis testing problem against independence over a noisy channel. In particular, we provided exact asymptotic characterization of the type-II error exponent subject to a mutual information privacy constraint on the information source and a constant constraint on the type-I error probability. Our results imply that the asymptotic privacy-utility tradeoff cannot be increased by tolerating a larger type-I error probability, which is known as a \emph{strong converse} theorem. The strong converse theorems for several other important problems, including~\cite{gilani2019distributed,sreekumar2019hypothesis,ahlswede1986hypothesis}, are either established or recovered from our results. 

To better understand the privacy-utility tradeoff, one could develop novel techniques to obtain second-order asymptotic result~\cite[Chapter 2]{TanBook} for the problem, which reveals the \emph{non-asymptotic} fundamental limit. Such a result is more intuitive for practical situations where both the observation and communication are limited (i.e., $n$ and $k$ are both finite). It is also interesting to generalize our proof ideas to derive or strengthen the privacy-utility tradeoff for other hypothesis testing or communication problems, e.g.,~\cite{sreekumar2018privacy,sankar2013utility}. Furthermore, one can study the privacy-utility tradeoff for the Bayesian setting~\cite{li2015privacy} of the present problem where the utility is the decay rate of the average of type-I and type-II error probabilities. Finally, one can also generalize our results to other privacy measures, such as the differential privacy~\cite{dwork2008differential}, the R\'enyi divergence~\cite{van2014renyi,zhou2020multiple}, the maximal leakage~\cite{liao2017hypothesis2,issa2019operational} or the maximal $\alpha$-leakage~\cite{liao2018tunable}.

\appendix

\subsection{Proof of \eqref{needjust} and \eqref{needjust2}}
\label{proof:needjust}

Recall the definition of the set $\calJ(\calA,\calB,P_A)$. Under the high privacy limit where $L=\frac{\rho^2}{2}$ for arbitrary small $\rho$, the privacy mechanism $P_{Z|U}$ can be written as\footnote{Readers can refer to \cite{liao2017hypothesis} for details.}
\begin{align}
P_{Z|U}(z|u)=Q_Z(z)+\rho J(u,z),
\end{align}  
where $J(u,z)$ is the $z$-th element of $u$-th row of a matrix $\bJ\in\calJ(\calU,\calZ,P_U)$.

Thus, for each $z\in\calZ$, the induced marginal distribution $P_Z$ of $P_U$ and $P_{Z|U}$ satisfies
\begin{align}
P_Z(z)
&=Q_Z(z)\label{equalmarginalPZ}.
\end{align}
Using Euclidean information theory~\cite{borade2008,huang2015euclidean}, we have that
\begin{align}
\nn&I(P_U,P_{Z|U})\\*
&=\sum_uP_U(u)D(P_{Z|U=u}\|P_Z)\\
&\approx \frac{1}{2}\sum_u P_U(u)\sum_z\frac{(P_{Z|U}(z|u)-P_Z(z))^2}{P_Z(z)}\\
&\approx \frac{\rho^2}{2}\sum_u P_U(u)\sum_z\frac{J(u,z)^2}{Q_Z(z)}\label{useapprox}.
\end{align}

Recall the definition of $\barQ_W$ in \eqref{def:QW}. The induced distributions $P_W$ and $P_{W|V}$ of $P_Z$ and $P_{W|Z}$ satisfy that for any $(v,w)\in\calV\times\calW$,
\begin{align}
P_W(w)&=Q_W(w),
\end{align}
and
\begin{align}
\nn&P_{W|V}(w|v)\\*
&=\barQ_W(w)+\rho\sum_{u,z}P_{U|V}(u|v)P_{W|Z}(w|z)J(u,z).
\end{align}
Similar to \eqref{useapprox}, using the definition of $h(\bJ,\rho)$ in \eqref{def:hvw}, we have
\begin{align}
\nn&I(P_V,P_{W|V})\\*
&=\sum_vP_V(v)D(P_{W|V=v}\|P_W)\\
&\approx\frac{1}{2} \sum_vP_V(v)\sum_w\frac{(P_{W|V}(w|v)-\barQ_W(w))^2}{\barQ_W(w)}\\
&\approx h(\bJ,\rho).
\end{align}
The justification of \eqref{needjust} is completed by combining these approximations.

If we further assume that $\tau C(P_{Y|X})=\frac{\rho^2}{2}$, then the conditional probability $P_{W|Z}$ should satisfy 
\begin{align}
P_{W|Z}(w|z)=Q_W(w)+\rho \Theta(z,w)
\end{align}
for any $Q_W\in\calQ(\calW)$, where $\Theta\in\calJ(\calZ,\calW,Q_Z)$.

Then we have that induced marginal distribution $P_W$ of $P_Z$ and $P_{W|Z}$ satisfies that for any $w\in\calW$,
\begin{align}
P_W(w)=Q_W(w).
\end{align}
Similar to \eqref{useapprox}, we have
\begin{align}
I(P_Z,P_{W|Z})
&=\sum_z P_Z(z)\log\frac{P_{W|Z}(w|z)}{P_W(w)}\\
&\approx \frac{\rho^2}{2}\sum_z P_Z(z)\sum_w\frac{(\Theta(z,w))^2}{Q_W(w)}\\
&=\frac{\rho^2}{2}\sum_{z,w} Q_Z(z)\frac{(\Theta(z,w))^2}{Q_W(w)}\label{usePz=Qz},
\end{align}
where \eqref{usePz=Qz} follows from \eqref{equalmarginalPZ}. The induced distributions $P_{W|V}$ and $P_W$ satisfy
\begin{align}
\nn&P_{W|V}(w|v)\\*
&=Q_W(w)+\rho^2\sum_{u,v}P_{U|V}(u|v)J(u,z)\Theta(z,w).
\end{align}
Similar to \eqref{useapprox}, we have
\begin{align}
\nn&I(P_V,P_{W|V})\\*
\nn&\approx \frac{\rho^4}{2}\sum_{v,w}\frac{P_V(v)}{Q_W(w)}\Big(\sum_{u,z}P_{U|V}(u|v)J(u,z)\Theta(z,w)\Big)^2.
\end{align}
The justification of \eqref{needjust2} is completed by combining above approximations for mutual information terms.

\subsection{Proof of Lemma \ref{alt:expression}}
\label{proof:alt:expression}
\subsubsection{Proof of Claim 1)}
From the definition of $g_{\lambda_1,\lambda_2}^{\tau,L}(\cdot)$ in \eqref{def:rb}, we have
\begin{align}
\nn&g_{\lambda_1,\lambda_2}^{\tau,L}(P_{UV},P_{Z|U},P_{Y|X})\\*
&=\sup_{\substack{Q_{UVZW}\in\calQ(P_{UV},P_{Z|U})\\Q_{XY}\in\calC:Q_{Y|X}=P_{Y|X}}}R_{\lambda_1,\lambda_2}^{\tau,L}(Q_{UVZW},Q_{XY})\\
\nn&=\sup_{Q_{UVZW}\in\calQ(P_{UV},P_{Z|U})}\Big(I(Q_V,Q_{W|V})-\lambda_1 I(Q_Z,Q_{W|Z})\\*
\nn&\qquad\qquad\qquad\qquad\qquad-\lambda_2 I(Q_U,Q_{Z|U})+\lambda_2 L\Big)\\*
&\qquad+\sup_{Q_{XY}\in\calC:Q_{Y|X}=P_{Y|X}}\lambda_1 \tau I(Q_X,Q_{Y|X}))\\
\nn&=\sup_{Q_{UVZW}\in\calQ(P_{UV},P_{Z|U})}\Big(I(Q_V,Q_{W|V})-\lambda_1 I(Q_Z,Q_{W|Z})\\*
&\qquad\qquad-\lambda_2 I(Q_U,Q_{Z|U})+\lambda_2 L\Big)+\lambda_1 \tau C(P_{Y|X})\label{use:capacity}\\
\nn&=\sup_{Q_{UVZW}\in\calQ(P_{UV},P_{Z|U})}\Big(I(Q_V,Q_{W|V})+\lambda_1\tau C(P_{Y|X})\\*
&\qquad-\lambda_1 I(Q_Z,Q_{W|Z})+\lambda_2(L-I(Q_U,Q_{Z|U}))\Big)\label{alt:rab},
\end{align}
where \eqref{use:capacity} follows from the definition of $C(P_{Y|X})$ in \eqref{def:capacity}.

On the one hand,
\begin{align}
\nn&g_{\lambda_1,\lambda_2}^{\tau,L}(P_{UV},P_{Z|U},P_{Y|X})\\*
&\geq\sup_{\substack{Q_{UVZW}\in\calQ(P_{UV},P_{Z|U}):\\I(Q_Z,Q_{W|Z})\leq \tau C(P_{Y|X})\\ I(Q_U,Q_{Z|U})\leq L}}I(Q_V,Q_{W|V})\\
&=f(\tau,L,P_{UV},P_{Z|U},P_{Y|X})\label{tousecritical}.
\end{align}
We then prove the other direction. For this purpose, let 
\begin{align}
\calR
\nn&:=\bigcup_{Q_{UVZW}\in\calQ(P_{UV},P_{Z|U})}\Big\{(\barE,\barR,\barL)\in\bbR_+^3:\\*
\nn&\qquad\barE\leq I(Q_V,Q_{W|V}),~\tau \barR\geq I(Q_Z,Q_{W|Z})\\*
&\qquad \barL\geq I(Q_U,Q_{Z|U})\Big\}\label{addlable2}.
\end{align}
It then follows that
\begin{align}
\nn&f(\tau,L,P_{UV},P_{Z|U},P_{Y|X})\\*
&=\sup\{\barE\in\bbR_+:~(\barE,C(P_{Y|X}),L)\in\calR\}\label{alt:f1}.
\end{align}

Consider any sequence of positive real numbers $\{\hatE_n\}_{n\in\bbN}$ such that
\begin{align}
\hatE_n=f(\tau,L,P_{UV},P_{Z|U},P_{Y|X})+\frac{1}{n}\label{def:En}.
\end{align}
Since $\hatE_n>f(\tau,L,P_{UV},P_{Z|U},P_{Y|X})$, from \eqref{alt:f1}, we have that $(\hatE_n,C(P_{Y|X}),L)\notin\calR$. Note that $\calR$ is a closed convex set. Applying the separating hyperplane theorem (cf.~\cite[Example 2.20]{boyd2004convex}), we conclude that there exists $(\lambda_1^*,\lambda_2^*)\in\bbR_+^2$ such that for any $(\barE,\barR,\barL)\in\calR$, 
\begin{align}
\hatE_n-\lambda_1^* \tau C(P_{Y|X})-\lambda_2^*L\geq \barE-\lambda_1^* \tau \barR-\lambda_2^*\barL\label{addalable}.
\end{align}
Thus, using \eqref{addlable2} and \eqref{addalable}, similarly to \cite{oohama2015wak,oohama2016wynerziv,zhou2016cilossy}, we have
\begin{align}
\nn&\hatE_n-\lambda_1^*\tau C(P_{Y|X})-\lambda_2^*L\\*
&\geq \sup_{(\barE,\barR,\barL)\in\calR}(\barE-\lambda_1^* \tau \barR-\lambda_2^* \barL)\\
\nn&\geq \sup_{Q_{UVZW}\in\calQ(P_{UV},P_{Z|U})} \Big(I(Q_V,Q_{W|V})-\lambda_1^* I(Q_Z,Q_{W|Z})\\*
&\qquad\qquad\qquad-\lambda_2^* I(Q_U,Q_{Z|U})\Big)\label{theotherdirec}.
\end{align}
Combining \eqref{alt:rab} and \eqref{theotherdirec} leads to 
\begin{align}
\hatE_n\geq g_{\lambda_1^*,\lambda_2^*}^{\tau,L}(P_{UV},P_{Z|U},P_{Y|X})\label{touse01}.
\end{align}
Using the definition of $E_n$ in \eqref{def:En}, we have
\begin{align}
\nn&f(\tau,L,P_{UV},P_{Z|U},P_{Y|X})\\*
&\geq g_{\lambda_1^*,\lambda_2^*}^{\tau,L}(P_{UV},P_{Z|U},P_{Y|X})-\frac{1}{n}\\
&\geq \min_{(\lambda_1,\lambda_2)\in\bbR_+^2}g_{\lambda_1,\lambda_2}^{\tau,L}(P_{UV},P_{Z|U},P_{Y|X})-\frac{1}{n}.
\end{align}
The proof is completed by taking $n\to\infty$.

\subsubsection{Proof of Claim 2)}
The definition of $g_{\lambda_1,\lambda_2}^{\tau,L}(\cdot)$ in \eqref{def:rb} implies
\begin{align}
\nn&g_{\lambda_1,\lambda_2}^{\tau,L}(P_{UV},P_{Z|U},P_{Y|X})\\*
&=\sup_{\substack{Q_{UVZW}\in\calQ(P_{UV},P_{Z|U})\\Q_{XY}\in\calC:Q_{Y|X}=P_{Y|X}}}R_{\lambda_1,\lambda_2}^{\tau,L}(Q_{UVZW},Q_{XY})\\
\nn&=\sup_{\substack{Q_{UVZW}\in\calQ(P_{UV},P_{Z|U})\\Q_{XY}\in\calC:Q_{Y|X}=P_{Y|X}}}\Big(R_{\lambda_1,\lambda_2}^{\tau,L}(Q_{UVZW},Q_{XY})\\*
&\qquad\quad-\Delta_{\gamma}^{\tau,L}(Q_{UVZW},Q_{XY},P_{UV},P_{Z|U},P_{Y|X})\Big)\label{useDelta}\\
\nn&\leq \sup_{\substack{Q_{UVZW}\in\calQ\\Q_{XY}\in\calC}}\Big(R_{\lambda_1,\lambda_2}^{\tau,L}(Q_{UVZW},Q_{XY})\\*
&\qquad\quad-\Delta_{\gamma}^{\tau,L}(Q_{UVZW},Q_{XY},P_{UV},P_{Z|U},P_{Y|X})\Big)\label{subset}\\
&=g_{\lambda_1,\lambda_2,\gamma}^{\tau,L}(P_{UV},P_{Z|U},P_{Y|X})\label{ub1},
\end{align}
where \eqref{useDelta} follows since $\Delta_{\gamma}^{\tau,L}(Q_{UVZW},Q_{XY},P_{UV},P_{Z|U},P_{Y|X})=0$ (cf. \eqref{def:Deltabg}) for $Q_{UVZW}\in\calQ(P_{UV},P_{Z|U})$ and $Q_{XY}\in\calC:Q_{Y|X}=P_{Y|X}$, \eqref{subset} follows since $\calQ(P_{UV},P_{Z|U})\subset\calQ$ and \eqref{ub1} follows from the definition of $g_{\lambda_1,\lambda_2,\gamma}^{\tau,L}(\cdot)$ in \eqref{def:rbg}.

For any $(\lambda_1,\lambda_2,\gamma)\in\bbR_+^3$, let $(Q_{UVZW}^{\lambda_1,\lambda_2,\gamma},Q_{XY}^{\lambda_1,\lambda_2,\gamma})$ be an optimizer of $g_{\lambda_1,\lambda_2,\gamma}^{\tau,L}(P_{UV},P_{Z|U},P_{Y|X})$ and let $Q_{\cdot}^{\lambda_1,\lambda_2,\gamma}$ be a distribution induced by either $Q_{UVZW}^{\lambda_1,\lambda_2,\gamma}$ or $Q_{XY}^{\lambda_1,\lambda_2,\gamma}$. From the support lemma~\cite[Appendix C]{el2011network}, we obtain that the cardinality of $W$ can be upper bounded as a function of $|\calU|$, $|\calV|$ and $|\calZ|$, which is finite. Furthermore, let $P_{UVZW}^{\lambda_1,\lambda_2,\gamma}$ and $P_{XY}^{\lambda_1,\lambda_2,\gamma}$ be defined as follows:
\begin{align}
P_{UVZW}^{\lambda_1,\lambda_2,\gamma}&=P_{UV}P_{Z|U}Q_{W|Z}^{\lambda_1,\lambda_2,\gamma}\label{def:uvzw},\\
P_{XY}^{\lambda_1,\lambda_2,\gamma}&=Q_X^{\lambda_1,\lambda_2,\gamma}P_{Y|X}.
\end{align}

Since KL divergence terms are non-negative~\cite{cover2012elements}, for any $(\lambda_1,\lambda_2)\in\bbR_+^2$,
\begin{align}
\nn&g_{\lambda_1,\lambda_2,\gamma}^{\tau,L}(P_{UV},P_{Z|U},P_{Y|X})\\*
\nn&=R_{\lambda_1,\lambda_2}^{\tau,L}(Q_{UVZW}^{\lambda_1,\lambda_2,\gamma},Q_{XY}^{\lambda_1,\lambda_2,\gamma})\\*
&\qquad-\Delta_{\gamma}^{\tau,L}(Q_{UVZW}^{\lambda_1,\lambda_2,\gamma},Q_{XY}^{\lambda_1,\lambda_2,\gamma},P_{UV},P_{Z|U},P_{Y|X})\\
&\leq R_{\lambda_1,\lambda_2}^{\tau,L}(Q_{UVZW}^{\lambda_1,\lambda_2,\gamma},Q_{XY}^{\lambda_1,\lambda_2,\gamma})\\
&\leq R_{\lambda_1,\lambda_2}^{\tau,L}(P_{UVZW}^{\lambda_1,\lambda_2,\gamma},P_{XY}^{\lambda_1,\lambda_2,\gamma})+\zeta(\lambda_1,\lambda_2,\gamma,\tau)\label{laststep}\\
&\leq g_{\lambda_1,\lambda_2}^{\tau,L}(P_{UV},P_{Z|U},P_{Y|X})+\zeta(\lambda_1,\lambda_2,\gamma,\tau)\label{laststep2},
\end{align}
where \eqref{laststep} is justified in Appendix \ref{just:laststep} and \eqref{laststep2} follows since $P_{UVZW}^{\lambda_1,\lambda_2,\gamma}\in\calQ(P_{UV},P_{Z|U})$ and $P_{XY}^{\lambda_1,\lambda_2,\gamma}$ satisfies that $P_{Y|X}^{\lambda_1,\lambda_2,\gamma}=P_{Y|X}$.

\subsection{Justification of \eqref{laststep}}
\label{just:laststep}
Note that $P_{UVZW}^{\lambda_1,\lambda_2,\gamma}$ in \eqref{def:uvzw} can be written equivalently as
\begin{align}
P_{UVZW}^{\lambda_1,\lambda_2,\gamma}
=P_ZQ_{W|Z}^{\lambda_1,\lambda_2,\gamma}P_{U|Z}P_{V|U}.
\end{align}
From the definitions of $(Q_{UVZW}^{\lambda_1,\lambda_2,\gamma},Q_{XY}^{\lambda_1,\lambda_2,\gamma})$ and $(P_{UVZW}^{\lambda_1,\lambda_2,\gamma},P_{XY}^{\lambda_1,\lambda_2,\gamma})$, we have
\begin{align}
D(Q_{UVZW}^{\lambda_1,\lambda_2,\gamma}\|P_{UVZW}^{\lambda_1,\lambda_2,\gamma})
\nn&=D(Q_Z^{\lambda_1,\lambda_2,\gamma}\|P_Z)\\*
&\!\!\!\!\!\!\!\!\!\!\!\!\!\!\!+D(Q_{UV|ZW}^{\lambda_1,\lambda_2,\gamma}\|P_{U|Z}P_{V|U}|Q_{ZW}^{\lambda_1,\lambda_2,\gamma}),\\
D(Q_{XY}^{\lambda_1,\lambda_2,\gamma}\|P_{XY}^{\lambda_1,\lambda_2,\gamma})
&=D(Q_{Y|X}^{\lambda_1,\lambda_2,\gamma}\|P_{Y|X}|Q_X^\gamma).
\end{align}

Furthermore,
\begin{align}
\nn&\gamma D(Q_{UVZW}^{\lambda_1,\lambda_2,\gamma}\|P_{UVZW}^{\lambda_1,\lambda_2,\gamma})
+\tau\gamma D(Q_{XY}^{\lambda_1,\lambda_2,\gamma}\|P_{XY}^{\lambda_1,\lambda_2,\gamma})\\
&=\Delta_{\gamma}^{\tau,L}(Q_{UVZW}^{\lambda_1,\lambda_2,\gamma},Q_{XY}^{\lambda_1,\lambda_2,\gamma},P_{UV},P_{Z|U},P_{Y|X})\label{use:defDeltaha}\\
&=R_{\lambda_1,\lambda_2}^{\tau,L}(Q_{UVZW}^{\lambda_1,\lambda_2,\gamma},Q_{XY}^{\lambda_1,\lambda_2,\gamma})-g_{\lambda_1,\lambda_2,\gamma}^{\tau,L}(P_{UV},P_{Z|U},P_{Y|X})\label{usefactoptimal}\\
&\leq R_{\lambda_1,\lambda_2}^{\tau,L}(Q_{UVZW}^{\lambda_1,\lambda_2,\gamma},Q_{XY}^{\lambda_1,\lambda_2,\gamma})\label{usedefs:rbrbg}\\
&\leq \log|\calV|+(\lambda_1+\lambda_2)\log|\calZ|+\lambda_1 \tau\log|\calY|\\
&=c(\lambda_1,\lambda_2,\tau)\label{usec:def},
\end{align}
where \eqref{use:defDeltaha} follows from the definition of $\Delta_{\gamma}^{\tau,L}(\cdot)$ in \eqref{def:Deltabg}, \eqref{usefactoptimal} follows since $(Q_{UVZW}^{\lambda_1,\lambda_2,\gamma},Q_{XY}^{\lambda_1,\lambda_2,\gamma})$ is an optimizer for $g_{\lambda_1,\lambda_2,\gamma}^{\tau,L}(P_{UV},P_{Z|U},P_{Y|X})$ (cf. \eqref{def:rbg}), \eqref{usec:def} follows from the definition of $c(\lambda_1,\lambda_2,\tau)$ in \eqref{def:cabt}, and \eqref{usedefs:rbrbg} follows since
\begin{align}
\nn&g_{\lambda_1,\lambda_2,\gamma}^{\tau,L}(P_{UV},P_{Z|U},P_{Y|X})\\*
&\geq g_{\lambda_1,\lambda_2}^{\tau,L}(P_{UV},P_{Z|U},P_{Y|X})\label{useub1}\\
&\geq f(\tau,L,P_{UV},P_{Z|U},P_{Y|X})\label{usecritical}\\
&\geq 0,
\end{align}
where \eqref{useub1} follows from the result in \eqref{ub1} and \eqref{usecritical} follows from \eqref{tousecritical}.

Note that the result in \eqref{usec:def} implies that
\begin{align}
D(Q_{UVZW}^{\lambda_1,\lambda_2,\gamma}\|P_{UVZW}^{\lambda_1,\lambda_2,\gamma})&\leq \frac{c(\lambda_1,\lambda_2,\tau)}{\gamma}\label{r1:klub},\\
 D(Q_{XY}^{\lambda_1,\lambda_2,\gamma}\|P_{XY}^{\lambda_1,\lambda_2,\gamma})&\leq \frac{c(\lambda_1,\lambda_2,\tau)}{\tau\gamma}\label{r2:klub}.
\end{align}

Using \eqref{r1:klub}, Pinsker's inequality and data processing inequality for KL divergence, we have
\begin{align}
\|Q_{VW}^{\lambda_1,\lambda_2,\gamma}-P_{VW}^{\lambda_1,\lambda_2,\gamma}\|
&\leq \sqrt{2D(Q_{VW}^{\lambda_1,\lambda_2,\gamma}\|P_{VW}^{\lambda_1,\lambda_2,\gamma})}\\
&\leq \sqrt{2D(Q_{UVZW}^{\lambda_1,\lambda_2,\gamma}\|P_{UVZW}^{\lambda_1,\lambda_2,\gamma})}\\
&\leq \sqrt{\frac{2c(\lambda_1,\lambda_2,\tau)}{\gamma}}.
\end{align}
Using \cite[Lemma 2.2.7]{csiszar2011information}, we have
\begin{align}
\nn&\big|H(Q_{VW}^{\lambda_1,\lambda_2,\gamma})-H(P_{VW}^{\lambda_1,\lambda_2,\gamma})\big|\\*
&\leq \sqrt{\frac{2c(\lambda_1,\lambda_2,\tau)}{\gamma}}\log\frac{|\calV||\calW|}{\sqrt{\frac{2c(\lambda_1,\lambda_2,\tau)}{\gamma}}}.
\end{align}
We can obtain similar upper bounds for $\big|H(Q_V^{\lambda_1,\lambda_2,\gamma})-H(P_V^{\lambda_1,\lambda_2,\gamma})\big|$ and $\big|H(\barQ_W^{\lambda_1,\lambda_2,\gamma})-H(P_W^{\lambda_1,\lambda_2,\gamma})\big|$.
Therefore,
\begin{align}
\nn&\big|I(\barQ_W^{\lambda_1,\lambda_2,\gamma},Q_{V|W}^{\lambda_1,\lambda_2,\gamma})-I(P_W^{\lambda_1,\lambda_2,\gamma},P_{V|W}^{\lambda_1,\lambda_2,\gamma})\big|\\*
\nn&\leq \big|H(Q_V^{\lambda_1,\lambda_2,\gamma})-H(P_V^{\lambda_1,\lambda_2,\gamma})\big|+\big|H(\barQ_W^{\lambda_1,\lambda_2,\gamma})-H(P_W^{\lambda_1,\lambda_2,\gamma})\big|\\*
&\qquad+\big|H(Q_{VW}^{\lambda_1,\lambda_2,\gamma})-H(P_{VW}^{\lambda_1,\lambda_2,\gamma})\big|\\
&\leq 3\sqrt{\frac{2c(\lambda_1,\lambda_2,\tau)}{\gamma}}\log\frac{|\calW||\calV|}{\sqrt{\frac{2c(\lambda_1,\lambda_2,\tau)}{\gamma}}}\label{result1:proved}.
\end{align}
Similarly to \eqref{result1:proved}, we have
\begin{align}
\nn&\big|I(\barQ_W^{\lambda_1,\lambda_2,\gamma},Q_{Z|W}^{\lambda_1,\lambda_2,\gamma})-I(P_W^{\lambda_1,\lambda_2,\gamma},P_{Z|W}^{\lambda_1,\lambda_2,\gamma})\big|\\*
&\leq 3\sqrt{\frac{2c(\lambda_1,\lambda_2,\tau)}{\gamma}}\log\frac{|\calW||\calZ|}{\sqrt{\frac{2c(\lambda_1,\lambda_2,\tau)}{\gamma}}}\label{result2:p},\\
\nn&\big|I(Q_X^{\lambda_1,\lambda_2,\gamma},Q_{Y|X}^{\lambda_1,\lambda_2,\gamma})-I(P_X^{\lambda_1,\lambda_2,\gamma},P_{Y|X}^{\lambda_1,\lambda_2,\gamma})\big|\\*
&\leq 2\sqrt{\frac{2c(\lambda_1,\lambda_2,\tau)}{\tau\gamma}}\log\frac{|\calX||\calY|}{\sqrt{\frac{2c(\lambda_1,\lambda_2,\tau)}{\tau\gamma}}}\label{result:3p},\\
\nn&\big|I(Q_Z^{\lambda_1,\lambda_2,\gamma},Q_{U|Z}^{\lambda_1,\lambda_2,\gamma})-I(P_Z^{\lambda_1,\lambda_2,\gamma},P_{Z|U}^{\lambda_1,\lambda_2,\gamma})\big|\\*
&\leq 3\sqrt{\frac{2c(\lambda_1,\lambda_2,\tau)}{\gamma}}\log\frac{|\calU||\calZ|}{\sqrt{\frac{2c(\lambda_1,\lambda_2,\tau)}{\gamma}}},
\end{align}
where distributions $Q_{\cdot}^{\lambda_1,\lambda_2,\gamma}$ is induced by either $Q_{UVZW}^{\lambda_1,\lambda_2,\gamma}$ or $Q_{XY}^{\lambda_1,\lambda_2,\gamma}$ and similarly for distributions $P_{\cdot}^{\lambda_1,\lambda_2,\gamma}$.

The justification of \eqref{laststep} is completed by combing \eqref{result1:proved} to \eqref{result:3p} with the following triangle inequality
\begin{align}
\nn&\big|R_{\lambda_1,\lambda_2}^{\tau,L}(Q_{UVZW}^{\lambda_1,\lambda_2,\gamma},Q_{XY}^{\lambda_1,\lambda_2,\gamma})-R_{\lambda_1,\lambda_2}^{\tau,L}(P_{UVZW}^{\lambda_1,\lambda_2,\gamma},P_{XY}^{\lambda_1,\lambda_2,\gamma})
\big|\\*
\nn&\leq \big|I(\barQ_W^{\lambda_1,\lambda_2,\gamma},Q_{V|W}^{\lambda_1,\lambda_2,\gamma})-I(P_W^{\lambda_1,\lambda_2,\gamma},P_{V|W}^{\lambda_1,\lambda_2,\gamma})\big|\\*
\nn&\qquad+\lambda_1\big|I(\barQ_W^{\lambda_1,\lambda_2,\gamma},Q_{Z|W}^{\lambda_1,\lambda_2,\gamma})-I(P_W^{\lambda_1,\lambda_2,\gamma},P_{Z|W}^{\lambda_1,\lambda_2,\gamma})\big|\\*
\nn&\qquad+\lambda_1 \tau\big|I(Q_X^{\lambda_1,\lambda_2,\gamma},Q_{Y|X}^{\lambda_1,\lambda_2,\gamma})-I(P_X^{\lambda_1,\lambda_2,\gamma},P_{Y|X}^{\lambda_1,\lambda_2,\gamma})\big|\\*
&\qquad+\lambda_2\big|I(Q_Z^{\lambda_1,\lambda_2,\gamma},Q_{U|Z}^{\lambda_1,\lambda_2,\gamma})-I(P_Z^{\lambda_1,\lambda_2,\gamma},P_{Z|U}^{\lambda_1,\lambda_2,\gamma})\big|.
\end{align}

\subsection{Proof of Lemma \ref{single-letter:terms}}
\label{proof:single}

Similarly to \cite[Proposition 1]{tyagi2020strong}, we have
\begin{align}
\nn&H(\tilU^k)+D(P_{\tilU^k}\|P_U^k)\\*
&=kH(\tilU_J)+kD(P_{\tilU_J}\|P_U),\\
\nn&H(\tilZ^k,\tilV^k)+D(P_{\tilZ^k\tilV^k}\|P_{ZV}^k)\\*
&=k\big(H(\tilZ_{J},\tilV_{J})+D(P_{\tilZ_{J}\tilV_{J}}\|P_{ZV})\big),\label{touse1}\\
\nn&H(\tilY^n|\tilX^n)+D(P_{\tilY^n|\tilX^n}\|P_{Y|X}^n|P_{\tilX^n})\\*
&=nH(\tilY_{J_n}|\tilX_{J_n})+nD(P_{\tilY_{J_n}|\tilX_{J_n}}\|P_{Y|X}|P_{X_{J_n}})\label{touse2}.
\end{align}

Then we have
\begin{align}
\nn&I(\tilU^k;\tilZ^k)\\*
&=H(\tilU^k)-H(\tilU^k|\tilZ^k)\\
\nn&=kH(\tilU_J)-\sum_{i\in[k]}H(\tilU_i|\tilU^{i-1},\tilZ^k)\\*
&\qquad+kD(P_{\tilU_J}\|P_U)-D(P_{\tilU^k}\|P_U^k)\\
&\geq kH(\tilU_J)-\sum_{i\in[k]}H(\tilU_i|\tilZ_i)-D(P_{\tilU^k}\|P_U^k)\\
&=kH(\tilU_J)-kH(\tilU_J|\tilZ_J)-D(P_{\tilU^k}\|P_U^k)\\
&=kI(\tilU_J;\tilZ_J)-D(P_{\tilU^k}\|P_U^k)\\
&\geq kI(\tilU_J;\tilZ_J)-2\log\frac{2}{1-\varepsilon}\label{useclose},
\end{align}
where \eqref{useclose} follows from the non-negativity of KL divergence and the result in \eqref{useresults}.

Furthermore, we have
\begin{align}
I(\tilY^n;\tilV^k)
&=\sum_{i\in[k]}I(\tilY^n;\tilV_i|\tilV^{i-1})\\
&\leq \sum_{i\in[k]}I(\tilV^{i-1},\tilY^n;\tilV_i)\\
&\leq \sum_{i\in[k]}I(\tilZ^{i-1},\tilV^{i-1},\tilY^n;\tilV_i)\\
&=\sum_{i\in[k]}I(W_i;\tilV_i)\\
&=kI(W_{J},J;\tilV_{J})\label{term1},
\end{align}
and
\begin{align}
\nn&I(\tilX^n;\tilY^n)\\*
&=H(\tilY^n)-H(\tilY^n|\tilX^n)\\
&=\sum_{i\in[n]}H(\tilY_i|\tilY^{i-1})-H(\tilY^n|\tilX^n)\\
&\leq \sum_{i\in[n]}H(\tilY_i)-H(\tilY^n|\tilX^n)\\
\nn&\leq nH(\tilY_{J_n})-nH(\tilY_{J_n}|\tilX_{J_n})\\*
&\qquad-nD(P_{\tilY_{J_n}|\tilX_{J_n}}\|P_{Y|X}|P_{X_J})+D(P_{\tilY^n|\tilX^n}\|P_{Y|X}^n|P_{\tilX^n})\label{use2}\\
&\leq nI(\tilX_{J_n};\tilY_{J_n},J_n)+D(P_{\tilZ^k\tilV^k\tilX^n\tilY^n}\|P_{Z^kV^kX^nY^n})\\
&\leq nI(\tilX_{J_n};\tilY_{J_n},J_n)-2\log(1-\varepsilon)+3\log 2\label{term22},
\end{align}
where \eqref{use2} follows from the result in \eqref{touse2} and \eqref{term22} follows from the result in \eqref{usedlsmall}. 

Similarly, we have
\begin{align}
\nn&I(\tilZ^k,\tilV^k;\tilY^n)\\*
&=H(\tilZ^k,\tilV^k)-H(\tilZ^k,\tilV^k|\tilY^n)\\
\nn&=kH(\tilZ_{J},\tilV_{J})+kD(P_{\tilZ_{J}\tilV_{J}}\|P_{ZV})-D(P_{\tilZ^k\tilV^k}\|P_{ZV}^k)\\*
&\qquad-\sum_{i\in[k]}H(\tilZ_i,\tilV_i|\tilZ^{i-1},\tilV^{i-1},\tilY^n)\label{use1}\\
&\geq kH(\tilZ_{J},\tilV_{J})+2\log(1-\varepsilon)+3\log 2-kH(\tilZ_{J},\tilV_{J}|W_J)
\label{usesmall2}\\
&=k I(\tilZ_{J},\tilV_{J};W_{J},J)+2\log(1-\varepsilon)+3\log 2\\
&\geq kI(\tilZ_{J};W_{J},J)+2\log(1-\varepsilon)+3\log 2\label{term21},
\end{align}
where \eqref{use1} follows from \eqref{touse1}, \eqref{usesmall2} follows similarly to  \eqref{usedlsmall}.

Furthermore, using non-negativity and convexity of KL divergence~\cite{cover2012elements}, we have
\begin{align}
\nn&D(P_{\tilU^k\tilZ^k\tilV^k\tilX^n\tilY^n}\|P_{U^kZ^kV^kX^nY^n})\\*
\nn&=D(P_{\tilZ^k}\|P_Z^k)+D(P_{\tilX^n|\tilZ^k}\|P_{X^n|Z^k}|P_{\tilZ^k})\\*
\nn&\qquad+D(P_{\tilY^n|\tilZ^k\tilX^n}\|P_{Y|X}^n|P_{\tilZ^k\tilX^n})\\*
&\qquad+D(P_{\tilU^k\tilV^k|\tilZ^k\tilX^n\tilY^n}\|P_{U|Z}^kP_{V|U}^k|P_{\tilZ^k\tilX^n\tilY^n})\\
\nn&\geq D(P_{\tilZ^k}\|P_Z^k)+D(P_{\tilY^n|\tilZ^k\tilX^n}\|P_{Y|X}^n|P_{\tilZ^k\tilX^n})\\*
&\qquad+D(P_{\tilU^k\tilV^k|\tilZ^k\tilX^n\tilY^n}\|P_{U|Z}^kP_{V|U}^k|P_{\tilZ^k\tilX^n\tilY^n})\\
\nn&\geq k D(P_{\tilZ_J}\|P_Z)+nD(P_{\tilY_{J_n}|\tilX_{J_n}}\|P_{Y|X}|P_{\tilX_{J_n}})\\*
&\qquad+k D(P_{\tilU_J\tilV_J|\tilZ_J\tilW_J}\|P_{U|Z}P_{V|U}|P_{\tilZ_J\tilW_J})\label{term4}.
\end{align}

\section*{Acknowledgments}
The authors would like to acknowledge four anonymous reviewers for many helpful comments and suggestions, which significantly improve the quality of the current manuscript. 

\bibliographystyle{IEEEtran}
\bibliography{IEEEfull_lin}

\begin{thebibliography}{10}
\providecommand{\url}[1]{#1}
\csname url@samestyle\endcsname
\providecommand{\newblock}{\relax}
\providecommand{\bibinfo}[2]{#2}
\providecommand{\BIBentrySTDinterwordspacing}{\spaceskip=0pt\relax}
\providecommand{\BIBentryALTinterwordstretchfactor}{4}
\providecommand{\BIBentryALTinterwordspacing}{\spaceskip=\fontdimen2\font plus
\BIBentryALTinterwordstretchfactor\fontdimen3\font minus
  \fontdimen4\font\relax}
\providecommand{\BIBforeignlanguage}[2]{{%
\expandafter\ifx\csname l@#1\endcsname\relax
\typeout{** WARNING: IEEEtran.bst: No hyphenation pattern has been}%
\typeout{** loaded for the language `#1'. Using the pattern for}%
\typeout{** the default language instead.}%
\else
\language=\csname l@#1\endcsname
\fi
#2}}
\providecommand{\BIBdecl}{\relax}
\BIBdecl

\bibitem{poor2013introduction}
H.~V. Poor, \emph{{An Introduction to Signal Detection and Estimation}}.\hskip
  1em plus 0.5em minus 0.4em\relax Springer Science \& Business Media, 1988.

\bibitem{chernoff1952measure}
H.~Chernoff, ``A measure of asymptotic efficiency for tests of a hypothesis
  based on the sum of observations,'' \emph{The Annals of Mathematical
  Statistics}, vol.~23, no.~4, pp. 493--507, 1952.

\bibitem{strassen1962asymptotische}
V.~Strassen, ``Asymptotische absch{\"a}tzungen in shannons
  informationstheorie,'' in \emph{Trans. Third Prague Conf. Information
  Theory}, 1962, pp. 689--723.

\bibitem{ahlswede1986hypothesis}
R.~Ahlswede and I.~Csisz{\'a}r, ``Hypothesis testing with communication
  constraints,'' \emph{IEEE Trans. Inf. Theory}, vol.~32, no.~4, pp. 533--542,
  1986.

\bibitem{Sohraby07}
K.~Sohraby, D.~Minoli, and T.~Znati, \emph{Wireless sensor networks:
  technology, protocols, and applications}.\hskip 1em plus 0.5em minus
  0.4em\relax John Wiley \& Sons, 2007.

\bibitem{sreekumar2019distributed}
S.~Sreekumar and D.~G{\"u}nd{\"u}z, ``Distributed hypothesis testing over
  discrete memoryless channels,'' \emph{IEEE Trans. Inf. Theory}, 2019.

\bibitem{sreekumar2020strong}
------, ``Strong converse for testing against independence over a noisy
  channel,'' in \emph{IEEE ISIT}, 2020.

\bibitem{csiszar2011information}
I.~Csisz\'ar and J.~K{\"o}rner, \emph{Information Theory: Coding Theorems for
  Discrete Memoryless Systems}.\hskip 1em plus 0.5em minus 0.4em\relax
  Cambridge University Press, 2011.

\bibitem{tyagi2020strong}
H.~{Tyagi} and S.~{Watanabe}, ``Strong converse using change of measure
  arguments,'' \emph{IEEE Trans. Inf. Theory}, vol.~66, no.~2, pp. 689--703,
  2020.

\bibitem{sankar2013utility}
L.~Sankar, S.~R. Rajagopalan, and H.~V. Poor, ``Utility-privacy tradeoffs in
  databases: An information-theoretic approach,'' \emph{IEEE Trans. Inf.
  Forensics Security}, vol.~8, no.~6, pp. 838--852, 2013.

\bibitem{liao2017hypothesis}
J.~Liao, L.~Sankar, V.~Y.~F. Tan, and F.~du~Pin~Calmon, ``Hypothesis testing
  under mutual information privacy constraints in the high privacy regime,''
  \emph{IEEE Trans. Inf. Forensics Security}, vol.~13, no.~4, pp. 1058--1071,
  2017.

\bibitem{liao2017hypothesis2}
J.~Liao, L.~Sankar, F.~P. Calmon, and V.~Y. Tan, ``Hypothesis testing under
  maximal leakage privacy constraints,'' in \emph{IEEE ISIT}.\hskip 1em plus
  0.5em minus 0.4em\relax IEEE, 2017, pp. 779--783.

\bibitem{gilani2019distributed}
A.~Gilani, S.~Belhadj~Amor, S.~Salehkalaibar, and V.~Y. Tan, ``Distributed
  hypothesis testing with privacy constraints,'' \emph{Entropy}, vol.~21,
  no.~5, p. 478, 2019.

\bibitem{sreekumar2018privacy}
S.~Sreekumar, A.~Cohen, and D.~G{\"u}nd{\"u}z, ``Privacy-aware distributed
  hypothesis testing,'' \emph{Entropy}, vol.~22, no.~6, p. 665, 2018.

\bibitem{zhou2020multiple}
L.~Zhou, ``Multiple private key generation for continuous memoryless sources
  with a helper,'' \emph{IEEE Trans. Inf. Forensics Security}, vol.~15, pp.
  2629--2640, 2020.

\bibitem{ye2005isit}
C.~Ye and P.~Narayan, ``The secret key~private key capacity region for three
  terminals,'' in \emph{IEEE ISIT}, 2005, pp. 2142--2146.

\bibitem{bloch2011physical}
M.~Bloch and J.~Barros, \emph{Physical-layer security: from information theory
  to security engineering}.\hskip 1em plus 0.5em minus 0.4em\relax Cambridge
  University Press, 2011.

\bibitem{csiszar2004secrecy}
I.~Csiszar and P.~Narayan, ``Secrecy capacities for multiple terminals,''
  \emph{IEEE Trans. Inf. Theory}, vol.~50, no.~12, pp. 3047--3061, 2004.

\bibitem{maurer1993}
U.~M. Maurer, ``Secret key agreement by public discussion from common
  information,'' \emph{IEEE Trans. Inf. Theory}, vol.~39, no.~3, pp. 733--742,
  1993.

\bibitem{viswanathan1997distributed}
R.~Viswanathan and P.~K. Varshney, ``Distributed detection with multiple
  sensors part i. fundamentals,'' \emph{Proceedings of the IEEE}, vol.~85,
  no.~1, pp. 54--63, 1997.

\bibitem{dwork2008differential}
C.~Dwork, ``Differential privacy: A survey of results,'' in \emph{International
  conference on theory and applications of models of computation}.\hskip 1em
  plus 0.5em minus 0.4em\relax Springer, 2008, pp. 1--19.

\bibitem{liao2018tunable}
J.~Liao, O.~Kosut, L.~Sankar, and F.~P. Calmon, ``A tunable measure for
  information leakage,'' in \emph{IEEE ISIT}, 2018, pp. 701--705.

\bibitem{dwork2011firm}
C.~Dwork, ``A firm foundation for private data analysis,'' \emph{Communications
  of the ACM}, vol.~54, no.~1, pp. 86--95, 2011.

\bibitem{du2012privacy}
F.~du~Pin~Calmon and N.~Fawaz, ``Privacy against statistical inference,'' in
  \emph{Proc. 50th Annu. Allerton Conf.}, 2012, pp. 1401--1408.

\bibitem{makhdoumi2014information}
A.~Makhdoumi, S.~Salamatian, N.~Fawaz, and M.~M{\'e}dard, ``From the
  information bottleneck to the privacy funnel,'' in \emph{IEEE ITW}, 2014, pp.
  501--505.

\bibitem{kalantari2018robust}
K.~Kalantari, L.~Sankar, and A.~D. Sarwate, ``Robust privacy-utility tradeoffs
  under differential privacy and hamming distortion,'' \emph{IEEE Trans. Inf.
  Forensics Security}, vol.~13, no.~11, pp. 2816--2830, 2018.

\bibitem{borade2008}
S.~Borade and L.~Zheng, ``Euclidean information theory,'' in \emph{IEEE IZS},
  2008, pp. 14--17.

\bibitem{huang2015euclidean}
S.-L. Huang, C.~Suh, and L.~Zheng, ``Euclidean information theory of
  networks,'' \emph{IEEE Trans. on Inf. Theory}, vol.~61, no.~12, pp.
  6795--6814, 2015.

\bibitem{lau2020ICCASP}
T.~S. {Lau} and W.~{Peng Tay}, ``Privacy-aware quickest change detection,'' in
  \emph{IEEE ICASSP}, 2020, pp. 5999--6003.

\bibitem{cover2012elements}
T.~M. Cover and J.~A. Thomas, \emph{Elements of information theory}.\hskip 1em
  plus 0.5em minus 0.4em\relax John Wiley \& Sons, 2012.

\bibitem{oohama2015exponent}
Y.~Oohama, ``Exponent function for one helper source coding problem at rates
  outside the rate region,'' in \emph{IEEE ISIT}, 2015, pp. 1575--1579.

\bibitem{oohama2016wynerziv}
------, ``Exponential strong converse for source coding with side information
  at the decoder,'' \emph{Entropy}, vol.~20, no.~5, p. 352, 2018.

\bibitem{zhou2016cilossy}
L.~Zhou, V.~Y.~F. Tan, and M.~Motani, ``Exponential strong converse for content
  identification with lossy recovery,'' \emph{IEEE Trans. Inf. Theory},
  vol.~64, no.~8, pp. 5879---5897, 2018.

\bibitem{liu2016brascamp}
J.~Liu, T.~A. Courtade, P.~Cuff, and S.~Verd{\'u}, ``Smoothing {Brascamp-Lieb}
  inequalities and strong converses for common randomness generation,'' in
  \emph{IEEE ISIT}, 2016, pp. 1043--1047.

\bibitem{wei2009strong}
W.~Gu and M.~Effros, ``A strong converse for a collection of network source
  coding problems,'' in \emph{IEEE ISIT}, 2009, pp. 2316--2320.

\bibitem{oohama2015wak}
Y.~Oohama, ``Exponential strong converse for one helper source coding
  problem,'' \emph{Entropy}, vol.~21, no.~6, p. 567, 2019.

\bibitem{boyd2004convex}
S.~Boyd and L.~Vandenberghe, \emph{Convex optimization}.\hskip 1em plus 0.5em
  minus 0.4em\relax Cambridge university press, 2004.

\bibitem{berry1941accuracy}
A.~C. Berry, ``The accuracy of the {Gaussian} approximation to the sum of
  independent variates,'' \emph{Transactions of the {A}merican mathematical
  society}, vol.~49, no.~1, pp. 122--136, 1941.

\bibitem{esseen1942liapounoff}
C.-G. Esseen, \emph{On the {Liapounoff} limit of error in the theory of
  probability}.\hskip 1em plus 0.5em minus 0.4em\relax Almqvist \& Wiksell,
  1942.

\bibitem{sreekumar2019hypothesis}
S.~Sreekumar and D.~G{\"u}nd{\"u}z, ``Hypothesis testing over a noisy
  channel,'' in \emph{IEEE ISIT}, 2019, pp. 2004--2008.

\bibitem{TanBook}
V.~Y.~F. Tan, ``Asymptotic estimates in information theory with non-vanishing
  error probabilities,'' \emph{{Foundations and Trends$\,$\textregistered $ $
  in Communications and Information Theory}}, vol.~11, no. 1--2, pp. 1--184,
  2014.

\bibitem{li2015privacy}
Z.~Li and T.~J. Oechtering, ``Privacy-aware distributed bayesian detection,''
  \emph{IEEE J. Sel. Topics Signal Process.}, vol.~9, no.~7, pp. 1345--1357,
  2015.

\bibitem{van2014renyi}
T.~Van~Erven and P.~Harremos, ``R{\'e}nyi divergence and {K}ullback-{L}eibler
  divergence,'' \emph{IEEE Trans. Inf. Theory}, vol.~60, no.~7, pp. 3797--3820,
  2014.

\bibitem{issa2019operational}
I.~Issa, A.~B. Wagner, and S.~Kamath, ``An operational approach to information
  leakage,'' \emph{IEEE Trans. Inf. Theory}, vol.~66, no.~3, pp. 1625--1657,
  2019.

\bibitem{el2011network}
A.~El~Gamal and Y.-H. Kim, \emph{Network Information Theory}.\hskip 1em plus
  0.5em minus 0.4em\relax Cambridge University Press, 2011.

\end{thebibliography}
\end{document}